\documentclass[ reprint, superscriptaddress, amsmath, amssymb, aps, prb ]{revtex4-1}

\usepackage{graphicx}
\usepackage{dcolumn}
\usepackage{bm}
\usepackage{amsmath}
\usepackage[
  colorlinks=true,
  urlcolor=blue,
  linkcolor=blue,
  citecolor=blue
]{hyperref}
\usepackage{epstopdf}
\usepackage{color}
\usepackage{float}
\begin{document}

\title{Structural and magnetic properties of the new quantum magnet BaCuTe$_2$O$_6$}

\author{A. Samartzis}
\altaffiliation{alexandros.samartzis@helmholtz-berlin.de}
\affiliation{\mbox{Helmholtz-Zentrum Berlin f\"{u}r Materialien und Energie GmbH, Hahn-Meitner Platz 1, D-14109 Berlin, Germany}}
\affiliation{\mbox{Institut f\"{u}r Festk\"{o}rperphysik, Technische Universit\"{a}t Berlin, Hardenbergstra{\ss}e 36, D-10623 Berlin, Germany}}

\author{D. Khalyavin}
\affiliation{\mbox{ISIS Facility, STFC Rutherford Appleton Laboratory, Oxfordshire OX11 0QX, UK}}

\author{A. T. M. N. Islam}
\affiliation{\mbox{Helmholtz-Zentrum Berlin f\"{u}r Materialien und Energie GmbH, Hahn-Meitner Platz 1, D-14109 Berlin, Germany}}

\author{S. Chillal}
\affiliation{\mbox{Helmholtz-Zentrum Berlin f\"{u}r Materialien und Energie GmbH, Hahn-Meitner Platz 1, D-14109 Berlin, Germany}}

\author{K. Siemensmeyer}
\affiliation{\mbox{Helmholtz-Zentrum Berlin f\"{u}r Materialien und Energie GmbH, Hahn-Meitner Platz 1, D-14109 Berlin, Germany}}

\author{K. Prokes}
\affiliation{\mbox{Helmholtz-Zentrum Berlin f\"{u}r Materialien und Energie GmbH, Hahn-Meitner Platz 1, D-14109 Berlin, Germany}}

\author{D. J. Voneshen}
\affiliation{\mbox{ISIS Facility, STFC Rutherford Appleton Laboratory, Oxfordshire OX11 0QX, UK}}
\affiliation{\mbox{Department of Physics, Royal Holloway University of London, Egham, TW20 0EX}}

\author{A. Senyshyn}
\affiliation{\mbox{Heinz Maier-Leibnitz Zentrum, Lichtenbergstra{\ss}e 1, D-85747 Garching, Germany}}

\author{B. Lake}
\altaffiliation{bella.lake@helmholtz-berlin.de}
\affiliation{\mbox{Helmholtz-Zentrum Berlin f\"{u}r Materialien und Energie GmbH, Hahn-Meitner Platz 1, D-14109 Berlin, Germany}}
\affiliation{\mbox{Institut f\"{u}r Festk\"{o}rperphysik, Technische Universit\"{a}t Berlin, Hardenbergstra{\ss}e 36, D-10623 Berlin, Germany}}
\date{\today}%

\begin{abstract}
We investigate the structural and magnetic properties of the new quantum magnet BaCuTe$_2$O$_6$. This compound is synthesized for the first time in powder and single crystal form. Synchrotron X-ray and neutron diffraction reveal a cubic crystal structure (P4$_1$32) where the magnetic Cu$^{2+}$ ions form a complex network. Physical properties measurements suggest the presence of antiferromagnetic interactions with a Curie-Weiss temperature of $\approx$~-33~K, while long-range magnetic order occurs at the much lower temperature of $\approx$~6.3~K. The magnetic structure, solved using neutron diffraction, reveals antiferromagnetic order along chains parallel to the a, b and c crystal axes. This is consistent with the magnetic excitations which resemble the multispinon continuum typical of the spin-1/2 Heisenberg antiferromagnetic chain. A consistent intrachain interaction value of $\approx$~34~K is achieved from the various techniques. Finally the magnetic structure provides evidence that the chains are coupled together in a non-colinear arrangement by a much weaker antiferromagnetic, frustrated hyperkagome interaction.
\begin{description}
\item[PACS numbers]
May be entered using the \verb+\pacs{#1}+ command.
\end{description}
\end{abstract}

\pacs{Valid PACS appear here}
\maketitle

\section{\label{sec:introduction}Introduction}

Quantum magnets are interesting because of their unusual ground states and excitations. They have strong quantum fluctuations which suppress long-range magnetic order and conventional spin-wave excitations, giving rise to novel states \cite{Balents2010, Lacroix}. Typically quantum magnets have low spin quantum number that enhances quantum fluctuations, and low dimensional or frustrated interactions between the magnetic ions that suppress static magnetism. The simplest example of a quantum magnet is the spin-1/2 antiferromagnetic chain where the magnetic ions are coupled in one dimension only by antiferromagnetic nearest neighbor interactions \cite{Cloizeaux1962, Lake2005, Lake2013}. This system does not develop long-range magnetic order even down to the lowest temperatures and its excitations are spinons which have fractional spin quantum number S=1/2 rather than spin-waves or magnons which have S=1 \cite{Mourigal2013}. In higher dimensions quantum fluctuations can still arise in the case of frustrated interactions between the magnetic ions such as due to triangular or tetrahedral motifs \cite{Bramwell2001, Gingras2014, Castelnovo2012, Gardner2010}. Famous examples are the kagome (two-dimensional (2D) network of corner-sharing triangles) and pyrochlore (three-dimensional (3D) network of corner-sharing tetrahedra) lattices \cite{Rau2018, Zhou2017}. In the case of extreme magnetic frustration a quantum spin liquid state may be realized where there is total suppression of static magnetism and the spins are moving coherently even at zero temperature \cite{Savary2017, Zhou2017, Fennell2014}.

Recently a new family of quantum magnets with chemical formula ACuTe$_2$O$_6$ (A=Sr, Pb) was reported \cite{Chillal2020, Koteswararao2014, Khuntia2016, Ahmed2015, Koteswararao2015, Koteswararao2016, Saeaun2020, Chillalarxiv}. These compounds have cubic space group P4$_1$32 where the magnetic Cu$^{2+}$ ions (S=1/2) occupy the 12d Wyckoff site. On its own the first neighbor interaction between the Cu$^{2+}$ ions couples them into isolated triangles, the second neighbor interaction couples them in a hyperkagome lattice (3D network of corner-sharing triangles) and the third neighbor interactions couples them into chains \cite{Chillal2020, Koteswararao2014, Ahmed2015}. Despite the fact that Sr$^{2+}$ and Pb$^{2+}$ have very similar sizes, SrCuTe$_2$O$_6$ and PbCuTe$_2$O$_6$ have very different interaction strengths and thus magnetic properties. In SrCuTe$_2$O$_6$ the third neighbor interaction is dominant and antiferromagnetic giving rise to spin-1/2 Heisenberg antiferromagnetic chains (HAFC), while the weaker first neighbor interactions act to couple these chains together and induce long-range magnetic order at a finite but suppressed N\'{e}el temperature (compared to the Curie-Weiss temperature) \cite{Ahmed2015, Koteswararao2015, Saeaun2020, Chillal2020}.  In contrast, for PbCuTe$_2$O$_6$ the first and second neighbor interactions are strong, antiferromagnetic and approximately equal to each other giving rise to a highly frustrated 3D network of corner-sharing triangles called the hyper-hyperkagome lattice (related to but distinctly different from the hyperkagome lattice), while the weaker third neighbor interactions act to reduce the degree of frustration \cite{Chillal2020, Koteswararao2014}. No long-range magnetic order has been observed in powder samples of PbCuTe$_2O_6$ down to 20~mK and the excitations form very broad diffusive spheres completely different from the sharp dispersive spin-waves of a conventional 3D magnet \cite{Khuntia2016}. These results suggest that PbCuTe$_2$O$_6$ is potentially close to a quantum spin liquid state.

In this paper, we investigate another member of the ACuTe$_2$O$_6$ family, namely BaCuTe$_2$O$_6$. It should be mentioned that BaCuTe$_2$O$_6$ has never been reported in the literature and has not been previously synthesized as far as we are aware. The Ba$^{2+}$ ion is larger than Sr$^{2+}$ and Pb$^{2+}$ thus we can expect that the exchange interactions and the magnetic properties may be different from those of SrCuTe$_2$O$_6$ and PbCuTe$_2$O$_6$. Here we report the first powder synthesis and growth of single crystals. We then determine the crystal structure using neutron and x-ray diffraction and find that BaCuTe$_2$O$_6$ is isostructural to other members of the family with cubic space group P4$_1$32. Next we investigate the magnetic properties using DC magnetic susceptibility and heat capacity which reveal the presence of a phase transition to long-range magnetic order. We then solve the magnetic structure using neutron diffraction and investigate the excitations using inelastic neutron scattering. Our proposed coupling scheme is compared to that of SrCuTe$_2$O$_6$ and PbCuTe$_2$O$_6$.

\section{\label{sec:Synthesis}Synthesis of B\lowercase{a}C\lowercase{u}T\lowercase{e}$_2$O$_6$}

\begin{figure}
\centering
\begin{tabular}{c @{\qquad} c }
\hspace*{-0.2cm}
\includegraphics[width=0.6\columnwidth,keepaspectratio]{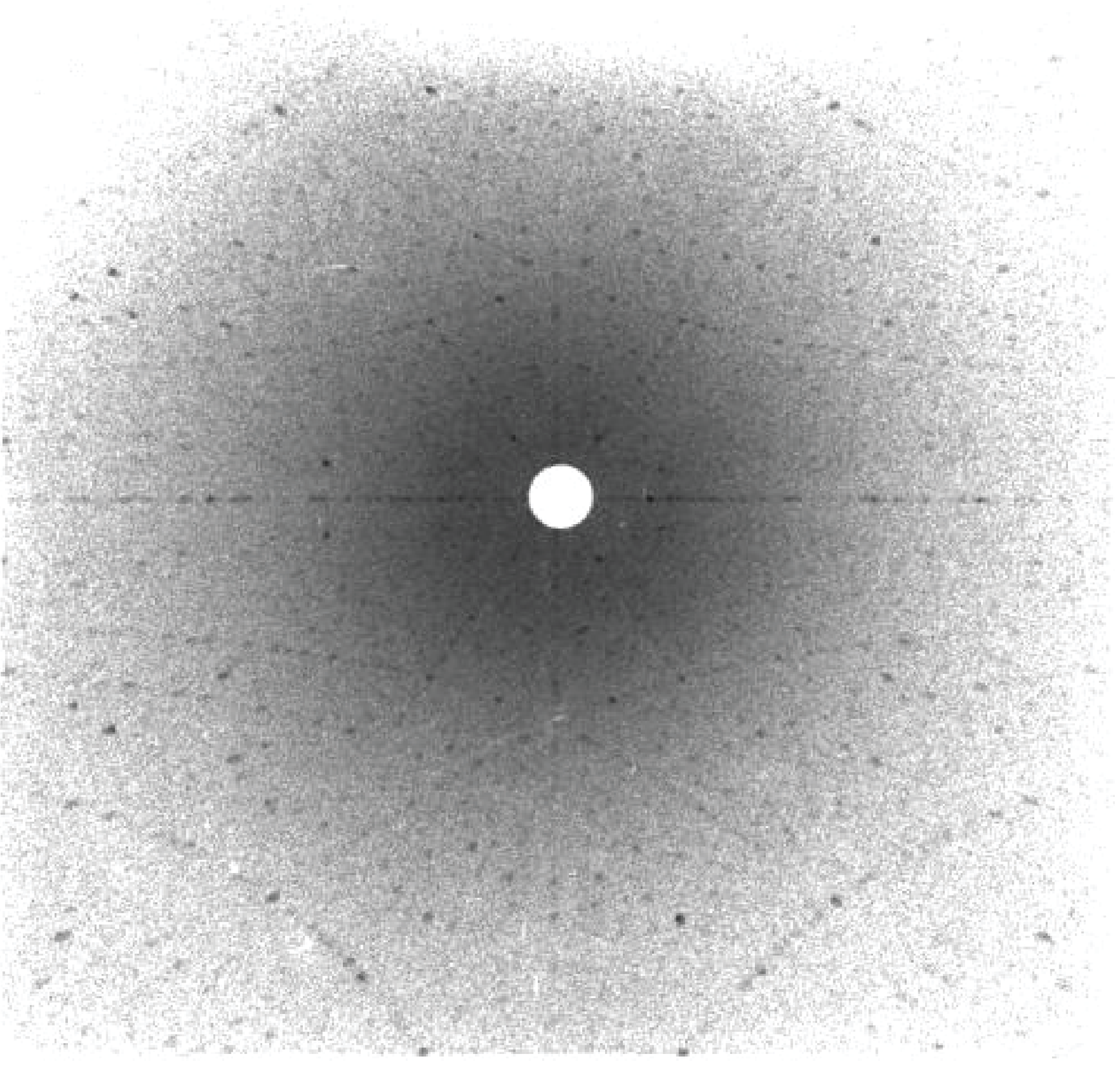} &
\hspace*{-4.0cm}
\includegraphics[width=1.2\columnwidth,keepaspectratio]{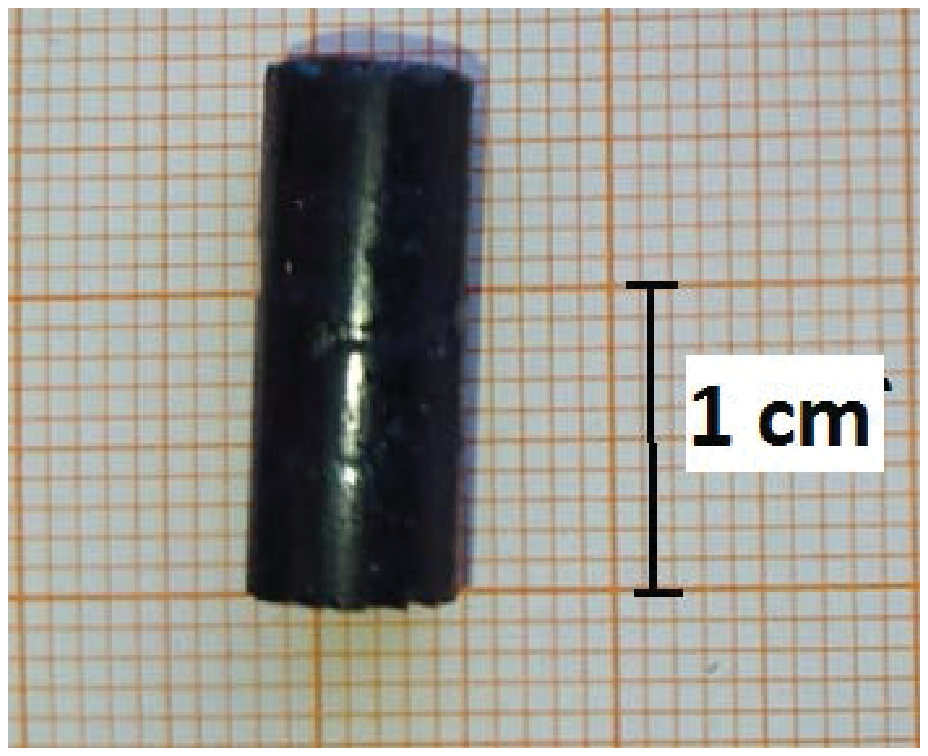}\\
\hspace*{-0.5cm}
\small (a) &
\hspace*{-4.0cm}
\small (b) 
\end{tabular}
\caption{(a) Laue diffraction pattern of a single crystal of BaCuTe$_2$O$_6$ with the X-ray beam parallel to the [1 1 0] direction. From the figure it can be deduced that the illuminated surface of the crystal is single phased. (b) Crystal III which was used for the inelastic neutron scattering measurements due to its size and quality.} 
\label{single_crystal}
\end{figure}

We have performed the first powder synthesis and single crystal growth of BaCuTe$_2$O$_6$. The polycrystalline sample was synthesized by solid-state reaction using high purity powders of  BaCO$_3$  (99.997\% Alfa Aesar), CuO (99.995\% Alfa Aesar) and TeO$_2$ (99.9995\% Alfa Aesar). The initial powders were mixed thoroughly in the 1:1:2 molar ratio and heated in a vacuum furnace twice for 12 hours at 650$^{\circ}$C under a flowing Argon atmosphere with intermediate grindings. In order to confirm the chemical composition of BaCuTe$_2$O$_6$, X-ray diffraction was performed at room temperature using a Brucker D8 machine which also revealed that the powder was of high quality.

The crystal growth was carried out using the stoichiometric powder described above. After the second firing the powder was pulverized, packed into a cylindrical rubber tube and pressed hydrostatically up to 2000 bars in a cold-isostatic-pressure (CIP) machine. The pressed rod was sintered in flowing Argon in a vacuum furnace for 12 hours at 650$^{\circ}$C to form a dense cylindrical rod with diameter $\approx$7~mm and length $\approx$7-8~cm. Crystal growth was performed in an optical Floating-zone furnace (Crystal Systems Corp., FZ-T 10000-H-VI-VPO) equipped with four 150~W Tungsten Halide lamps under 0.2~MPa Argon atmosphere at a growth rate of 1.0~mm/h. The single crystallinity of the as-grown crystal was checked by a back scattering X-ray (molybdenum tube) Laue diffractometer (see Fig.~\ref{single_crystal}(a)) and the cross-section was checked by Polarized Optical Microscopy. A piece of the crystal was also ground and checked with X-ray powder diffraction (Brucker D8) and found to be phase pure. From several growths, we identified three single crystals which were used for the present study, crystal III which has a length of 15.8~mm, diameter 5.5~mm and weight 2.884~g is shown in Fig.~\ref{single_crystal}(b). The synthesis and chacarterization of BaCuTe$_2$O$_6$ took place at the Core Laboratory for Quantum Materials, Helmholtz Zentrum Berlin f\"{u}r Materialien und Energie, Germany.

\section{\label{sec:methods}Experimental Details}

The crystal structure of the single crystals was studied via a synchrotron X-ray diffraction at room temperature, performed using a wavelength of 0.493~\AA\ on the Materials Science-beamline (MS-X04SA) \cite{MS-beamline} at the Paul Scherrer Institute (PSI) Switzerland. A small piece of single crystal was finely ground and mixed with diamond powder (ratio 1:3). The mixture was then placed in a thin capillary of 0.3~mm diameter which was spun continuously to avoid the effects of preferential orientation. The diffractometer measured a $2\theta$ range of 110$^{\circ}$ with a resolution 0.0036$^{\circ}$. 

Static dc-susceptibility ($\chi_{\text{dc}}$) measurements were performed on a powder sample pressed into a thin pellet ($\sim$17.06(5)~mg), over the temperature range 2-400~K under an applied magnetic field of 0.1~T. Heat capacity ($C_{\text{P}}$) measurements were performed on a pressed powder sample of $\sim$10.13(5)~mg. Data were collected in the temperature range 2-100~K for zero applied magnetic field. These measurements were performed on a Physical Properties Measurement System (PPMS, Quantum Design) and a Magnetic Properties Measurement System (MPMS, SQUID VSM, Quantum Design) located at the Core Laboratory for Quantum Materials, Helmholtz Zentrum Berlin f\"{u}r Materialien und Energie, Germany. 

Further structural characterization of BaCuTe$_2$O$_6$ was performed using the high resolution neutron powder diffractometer SPODI \cite{Hoelzel2012}, at the Heinz Maier-Leibnitz Zentrum (MLZ) in Munich, Germany. A powder sample of mass 11.1(1)~g was placed inside a copper can which was cooled in a $^{3}$He cryostat. The diffraction patterns at $T$=0.5 and 15~K were collected, using an incident wavelength of 1.548(2)~\AA\  and a detector coverage of $2\theta$=151$^{\circ}$ with a step size of 0.1$^{\circ}$. 

The magnetic structure of BaCuTe$_2$O$_6$ was refined from data \cite{doiWISH} collected on a powder sample of mass 11.1(1)~g measured by the high flux, high resolution neutron time-of-flight diffractometer (ToF) WISH \cite{Wish}, operated at Target Station 2 of the ISIS facility, Rutherford Appleton Laboratory, UK.  A solid-methane moderator transmits incident neutrons between 1.5-15~\AA. Diffraction patterns were collected using 10 detector banks covering a total $2\theta$-range of 160$^{\circ}$ and providing high resolution data over the whole wavevector-range. Data were collected between $T$=2 and 8~K in 1~K steps. The FullProf Suite \cite{Fullprof1993} was used to refine the diffraction data.

\begin{table}\label{AtomicCoord}
\footnotesize
\caption{\label{AtomicCoord}Atomic coordinates along with the thermal parameters, $B_{\text{iso}}$ from the Rietveld refinement within the P4$_1$32 space group of the X-ray synchrotron pattern at room temperature of the crushed single crystal (top section) and the SPODI neutron scattering pattern of the powder sample prepared by solid state reaction at 15~K (bottom section).} 
\begin{center}
\resizebox{\columnwidth}{!}{%
\begin{tabular}{ |c|c|c|c|c|c| }
\hline
\multicolumn{6}{|c|}{ } \\
\multicolumn{6}{|c|}{Synchrotron X-ray, crushed single crystal, 300 K} \\
\cline{1-6}
Atom & Mult & x & y & z & $B_{\text{iso}}$  \\
\hline
Te    & 24e& 0.3368(1) &0.9170(1) & 0.0641(1) & 1.14(1)  \\
 Ba1    & 8c  & 0.0586(1) & 0.0586(1) &0.0586(1) & 1.18(3)  \\
 Ba2    &4b & 0.3750 &0.6250 &  0.1250 &1.11(4)  \\
 Cu    &12d&  0.4741(1) &0.8750 &0.2758(1) & 1.18(5)  \\
 O1  &24e&     0.5794(6) &  0.9231(6) & 0.3697(6) & 1.42(18)  \\
 O2  &24e&     0.2658(5) & 0.8121(5) & 0.9923(6) & 1.05(18)  \\
 O3  &24e&  0.2227(6) &  0.9743(7) &  0.1342(7) &  2.66(21)  \\
\cline{1-6}
\multicolumn{6}{|c|}{ } \\
\multicolumn{6}{|c|}{Neutron, powder by solid state reaction, 15 K} \\
\cline{1-6}
Atom & Mult & x & y & z & $B_{\text{iso}}$  \\
\hline
  Te   & 24e &  0.3371(3) &  0.9180(3) &  0.0640(3) &  0.98(9) \\  
 Ba1    & 8c&   0.0563(3)&   0.0563(3)&   0.0563(3)&   0.36(19)  \\  
 Ba2  &4b   &   0.3750&   0.6250  & 0.1250 &  0.96(24)       \\
 Cu   &12d&  0.4738(2)  & 0.8750  & 0.2761(2)  & 0.80(10)    \\  
 O1   &24e&    0.5763(3) &  0.9242(3)&   0.3737(3)&   1.02(10) \\  
 O2  &24e &    0.2662(2) &  0.8115(5)&   0.9900(3)&   1.42(10) \\   
 O3   &24e&    0.2267(3) &  0.9739(3)&   0.1372(3)&   1.39(9)   \\
\cline{1-6}
\end{tabular}
}
\end{center}
\end{table}

The spin dynamics of BaCuTe$_2$O$_6$ were studied by inelastic neutron scattering (INS). Data for both a powder and single crystal sample \cite{doiLET} were collected on the LET \cite{LET}, cold neutron multi-chopper ToF spectrometer operating at Target Station 2 of the ISIS facility, Rutherford Appleton Laboratory, UK. The sample was cooled using an orange cryostat and data were measured at $T$=1.8~K and 8~K. The spectrum of the powder sample (mass 10.1(4)~g) was collected simultaneously using the five independent incident energies $E_\text{i}$= 14.1, 4.88, 2.45, 1.47 and 0.978~meV with corresponding elastic energy resolutions, $\Delta E_\text{i}$= 0.697, 0.144, 0.053, 0.025 and 0.014~meV for a total current of 240~$\mu$A and 310~$\mu$A at low and high temperatures respectively. The single crystal sample consisted of three co-aligned crystals (of total mass 4.80(1)~g) oriented with the (1,-1,0) crystallographic direction vertical.  The crystals were wrapped in Al foil and attached using Al wire to an Al sample holder. This sample was measured at $T$=1.8~K and $T$=8~K with incident energies, $E_{\text{i}}$= 13.75, 7.6, 4.81, 3.32 and 2.43~meV and corresponding elastic energy resolutions, $\Delta E_{\text{i}}$= 0.782, 0.333, 0.175, 0.105 and 0.068~meV. During the measurements the crystals were rotated over an angular range of 159$^{\circ}$ in steps of  0.5$^\circ$  or 1$^\circ$ for a current of 8$\mu$A per step.

\section{\label{sec:results}Results and discussions}

\subsection{Crystal Structure}

\begin{figure}
\begin{center}
\hspace*{-1.5cm}    
\includegraphics[width=1.5\columnwidth,keepaspectratio]{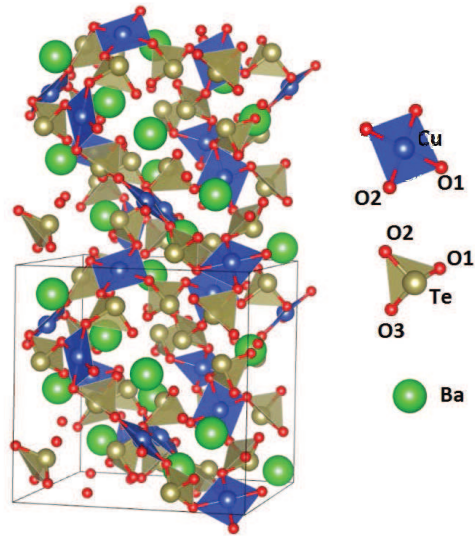}
\caption{Crystal structure of BaCuTe$_2$O$_6$ using the lattice paramaters and atomic coordinates extracted from the analysis of the high-resolution synchrotron powder X-ray diffraction data at room temperature. The ions Ba$^{2+}$, Cu$^{2+}$, Te$^{4+}$ and O$^{2-}$ are shown as green, blue, gold and red spheres respectively and the coordination of Te$^{4+}$ and Cu$^{2+}$ by O$^{2-}$ are shown by the polyhedra. The figure was prepared using the VESTA software \cite{VESTA}. } \label{CrystalStructure}
\end{center}
\end{figure}

\begin{figure}
\centering
\includegraphics[width=1\columnwidth,keepaspectratio]{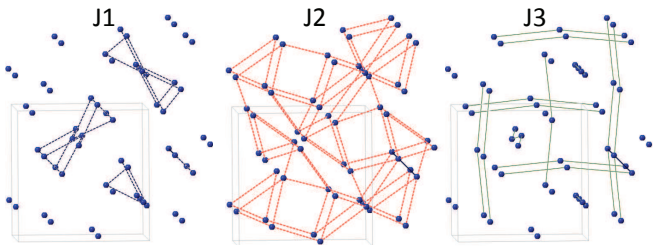}
\caption{The magnetic exchange interactions J1, J2 and J3 between the Cu$^{2+}$ ions in BaCuTe$_2$O$_6$ are presented from left to right. The first neighbor interactions, J1 (black dashed lines), form isolated Cu$^{2+}$ triangles. The second neighbor interactions, J2 (red dashed lines), lead to a 3D network of corner sharing triangles, known as the hyper-Kagome lattice. The third neighbor interactions J3, (green lines) form chains along the a-, b- and c-axes. Only the Cu$^{2+}$ ions are shown (blue spheres) and the light gray lines indicate the boundaries of one unit cell. The figure was prepared by the Crystal Maker software \cite{CrystalMaker}. }
\label{ExchangeInteractions}
\end{figure}

BaCuTe$_2$O$_6$ is a new compound that, to our knowldge, has never been previously reported in the literature, our first task was therefore to determine its crystal structure. We refined both synchrotron powder X-ray diffraction at 300~K measured on a crushed single crystal, and neutron diffraction collected from the powder sample on two different instruments at several temperature from 0.5~K to 15~K. The cubic space group $P$4$_1$32 which describes the related compounds SrCuTe$_2$O$_6$ and PbCuTe$_2$O$_6$ was found to account entirely for the crystal structure of BaCuTe$_2$O$_6$ at all temperatures. At room temperature the lattice parameter was a=12.8330(2)~\AA\ (see Table~\ref{CP}) and the atomic coordinates are listed in Table~\ref{AtomicCoord} along with the thermal parameters, $B_{iso}$. The lattice parameter decreases slightly with decreasing temperature but no significant change in the atomic positions are found at low temperatures. Details of this analysis along with selected fitted diffraction patterns are given in Appendix~\ref{sec:StructDet}. 

\begin{table}\label{CP}
\footnotesize
\caption{\label{CP}Crystallographic parameters of BaCuTe$_2$O$_6$ single crystal and powder samples obtained from the refinement of the synchrotron X-ray and neutron diffraction patterns at several temperatures. The lattice parameter a and refinement quality parameter, $R_{\text{Bragg}}$ are listed. The data were refined with the cubic space group $P$4$_1$32 ~($\#$213) which account entirely for the cystal structure at all temperatures. The refinement of the WISH powder data include two patterns from Banks 1 $\&$ 10 and Banks 2 $\&$ 9. The data at 2~K include a magnetic phase in the refinement.}
\begin{center}
\resizebox{\columnwidth}{!}{%
\begin{tabular}{ |c|c|c| }
\cline{1-3}
&&\\
~~~ $T$~(K) ~~~& ~~~~~~~~ a $(\AA)$ ~~~~~~~~ &  $R_{\text{Bragg}}$ ($\%)$ \\
&&\\
\hline
\multicolumn{3}{|c|}{Synchrotron X-rays, crushed single crystal} \\
\hline
300&12.8330(2)  & 9.87 \\
&&\\
\hline
\multicolumn{3}{|c|}{Neutrons @ SPODI, powder from solid state reaction} \\
\hline
15   & 12.8328(1) & 3.0\\
&&\\
0.5  & 12.8218(8) & 2.41\\
\hline
\multicolumn{3}{|c|}{Neutrons @ WISH, powder from solid state reaction} \\
\hline
8   & 12.8287(2) &2.53\\
&&\\
2  & 12.8277(8)  & 4.51\\
\cline{1-3}
\end{tabular}
}
\end{center}
\end{table}

The crystal structure of BaCuTe$_2$O$_6$ is presented in Fig.~\ref{CrystalStructure}. The Cu$^{2+}$ ions in BaCuTe$_2$O$_6$ are magnetic with spin S=1/2 angular momentum, there are 12 Cu$^{2+}$ ions in each unit cell ocupying the 12d Wyckoff site. Magnetic exchange interactions (super super-exchange interactions) are expected between neighboring Cu$^{2+}$ ions. Using the crystallographic parameters from the refinement of the synchrotron X-ray data, up to three nearest neighbor  Cu$^{2+}$- Cu$^{2+}$ bonds with successively increasing bond lengths are identified. The first nearest neighbor interaction, J1 (bond distance d1=4.73(8)~$\AA$), couples the Cu$^{2+}$ ions into isolated equilateral triangles. The second nearest neighbor interaction, J2 (bond distance d2=5.65(8)~$\AA$), forms a network of corner-sharing equilateral triangles known as the hyperkagome lattice. Finally the third neighbor interaction, J3 (bond distance d3=6.48(0)~$\AA$), connects the Cu$^{2+}$ ions into uniform chains along the a, b and c crystallographic directions. These interactions are presented individually in Fig.~\ref{ExchangeInteractions}.  It should be mentioned that each Cu$^{2+}$ ion is equivalent (single Wyckoff site 12d) and participates in one J1 triangle, two J2 triangles and one J3 chain. In the case of antiferromagnetic J1 and/or J2, geometrical magnetic frustration can be expected. The same set of interactions is found in the isostructural compounds PbCuTe$_2$O$_6$ and  SrCuTe$_2$O$_6$ \cite{Koteswararao2014, Koteswararao2016}.

\subsection{Bulk physical properties}

\subsubsection{Static magnetic susceptibility}\label{sec:magnetization}

\begin{figure}
\centering
\begin{tabular}{c @{\qquad} c }
\includegraphics[width=0.85\columnwidth,keepaspectratio]{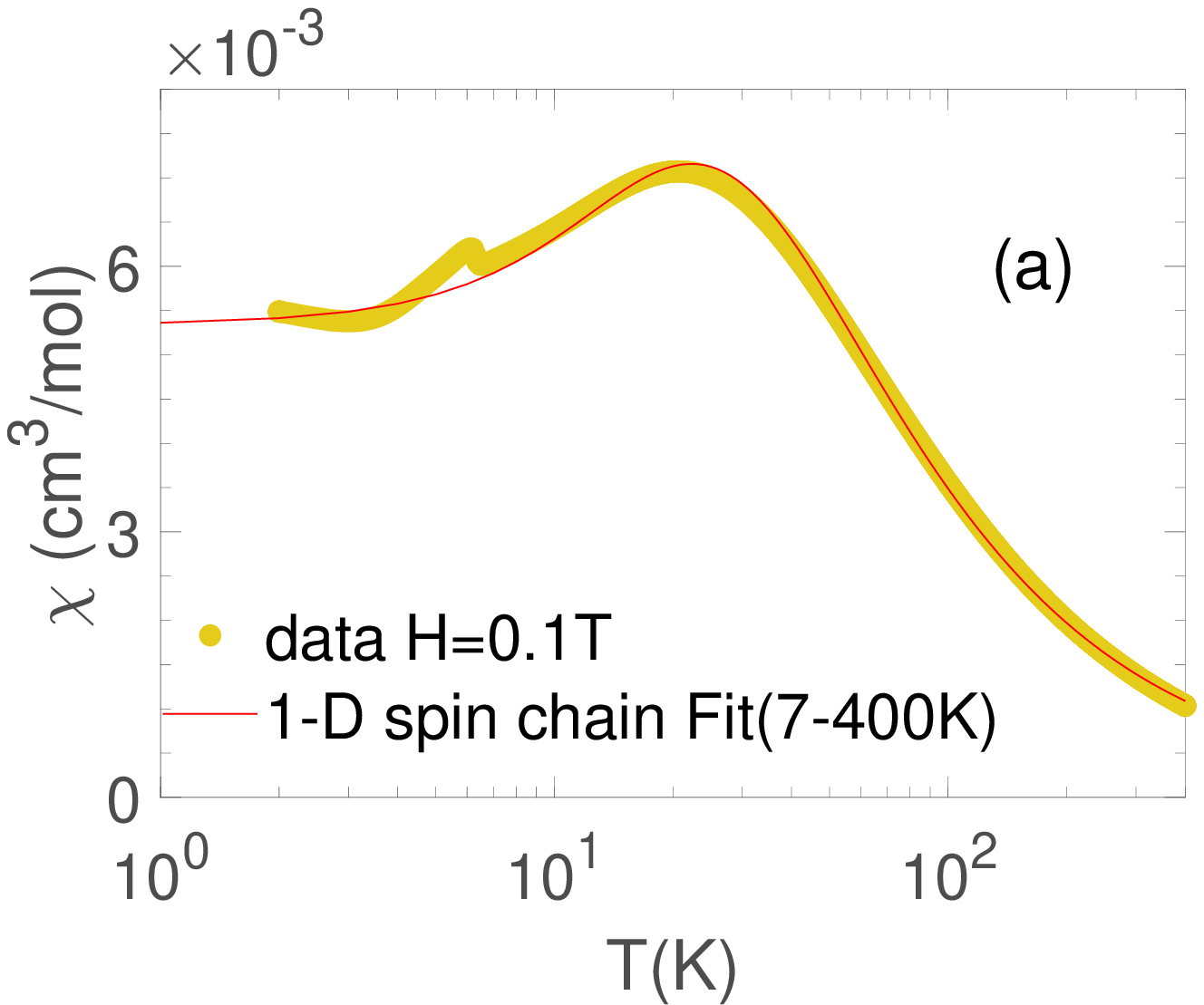} \\

\includegraphics[width=0.92\columnwidth,keepaspectratio]{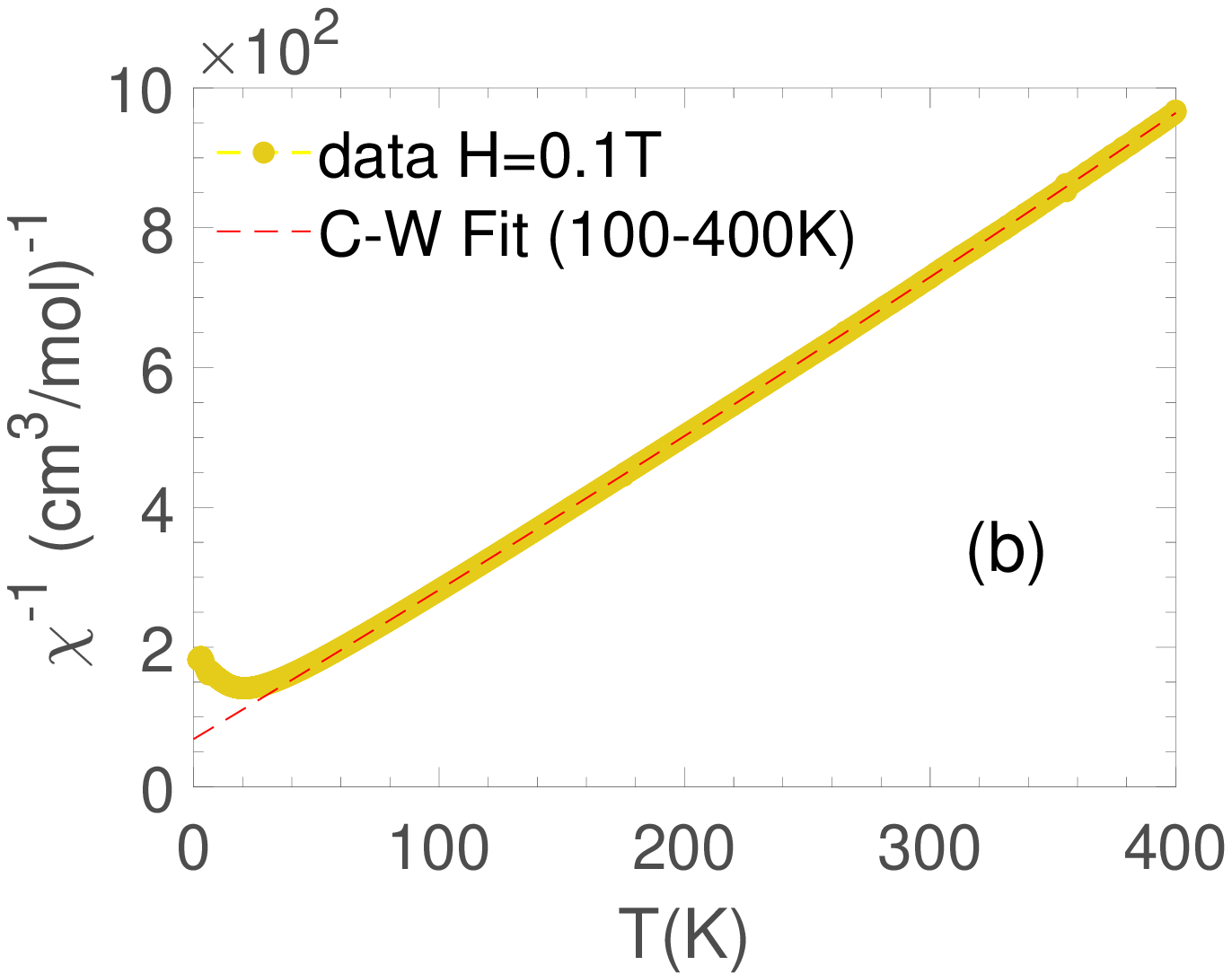}  \\

 \end{tabular}
\caption{(a) Temperature dependence of the static magnetic susceptibility, $\chi_{\text{dc}}$, measured on a powder sample of BaCuTe$_2$O$_6$ in an applied magnetic field $H$=0.1~T. The data in the range 7~K to 400~K were fitted by Eq.~\ref{1D} which includes the S-1/2 HAFC model. (b) Inverse susceptibility as a function of temperature. The data in the range 100~K to 400~K were fitted to the Curie-Weiss Law.}
\label{xdc_plot}
\end{figure}

In order to investigate the magnetic properties of BaCuTe$_2$O$_6$, static magnetic susceptibility data, $\chi_{\text{dc}}$, were collected on the powder sample, in the temperature range $T$=2 to 400~K for an applied magnetic field of $H$=0.1~T. Fig.~\ref{xdc_plot}(a) shows three distinctive temperature regions. The data at high temperature ($T>50$~K) decreases smoothly with increasing temperature suggesting a paramagnetic state, while a broad maximum at $\sim$22~K reflects the presence of short-range spin correlations which are a characteristic feature of magnets with low-dimensional and/or frustrated interactions. The prominent sharp peak at $T_\text{N}$=6.31~K reveals a magnetic phase transition to long-range magnetic order, while the susceptibility data increases with decreasing temperature below the transition suggesting a paramagnetic contribution from impurities and/or defects in the sample. 

The data in the range 100~K to 400~K were fitted by the modified Curie-Weiss expression $\chi(T)=\chi_0+C/(T+T_{\text{CW}})$ as shown by the red dotted line in Fig.~\ref{xdc_plot}(b) that presents the inverse susceptibility as a function of temperature. The constant term in the expression is associated with the susceptibility from core diamagnetism. $T_{\text{CW}}$ is the Curie-Weiss temperature and the parameter $C$ is the Curie constant that is related to the effective moment via the relation $\mu_{\text{eff}}$=$\sqrt{3k_\text{B}C/N_\text{A}}$ where $k_\text{B}$ and $N_\text{A}$ are the Boltzmann constant and the Avogadro number respectively. The best fit yields the values $\chi_0$~=~-0.0001(0)~cm$^3$/mol and $C$=0.4806(3)~cm$^3$K/mol from which the effective moment $\mu_{\text{eff}}$~=~1.96(5)~$\mu_{\text{B}}$/Cu$^{2+}$ was extracted. This value is slightly higher than the spin-only value of the free S=1/2 Cu$^{2+}$ ion ($\mu_{\text{eff}}$~=~ g$\sqrt{S(S+1)}$ $\mu_{\text{B}}\approx1.73$~$\mu_{\text{B}}$, for g=2). Finally the fitting suggests antiferromagnetic spin correlations with $T_{\text{CW}}$~=~-32.8(1)~K. In an unfrustrated magnet with 3D interactions the transition to long-range magnetic order occurs at a temperature close to the Curie-Weiss temperature, however in BaCuTe$_2$O$_6$  $T_\text{N}$=6.31~K is almost five times smaller than $T_{\text{CW}}$, this indicates the suppression of static magnetic order due magnetic frustration and/or low-dimensional interactions. 
  
In order to acquire insight into the broad maximum at $T\approx22$~K, the susceptibility data were compared to a model for the spin-1/2 Heisenberg antiferromagnetic chain (HAFC). This model was selected because of the similarity of the susceptibility of BaCuTe$_2$O$_6$ with that of the isostructural compound SrCuTe$_2$O$_6$, where the J3 interaction is dominant and couples the Cu$^{2+}$ ions into antiferromagnetic chains along the a, b and c crystallographic directions. We used the expression \cite{Johnston2000} 
\begin{equation}
\label{1D}
\chi(T)~=~\chi_0~+\chi_{\text{1D}}(T)
\end{equation} 
where the $\chi_{\text{1D}}(T)$ term is an expression for the susceptibility of the spin-1/2 HAFC calculated by S.~Eggert {\it et. al.} \cite{Eggert1994}
\begin{equation}
\label{1-d chain 1}
\chi_{\text{1D}}(T)=g^2(\frac{N_{\text{A}}\mu_{\text{B}}^2}{4k_{\text{B}}})F\left(\frac{J3}{k_{\text{B}}T}\right)\frac{1}{T} \\
\end{equation} 

where
\begin{equation}
\label{1-d chain 2}
F(x)=\frac{1+0.08516x+0.23351x^2}{1+0.73382x+0.13696x^2+0.53568x^3} \\
\end{equation} 

The $\chi_{\text{dc}}$ data in Fig.~\ref{xdc_plot}(a) were fitted by Eq.~\ref{1D} (red line) in the temperature range 7 to 400~K. The $\chi_{\text{1D}}$ term yields $g\approx2.096\pm0.002$ and J3$/k_{\text{B}}$=$34.40(4)$~K for the g-factor and intrachain interaction strength respectively, while the constant term is $\chi_0$=0.0001(0)~cm$^3$/mol. The effective moment, given by $\mu_{\text{eff}}$=$\sqrt{3k_\text{B}C_{\text{1D}}/N_\text{A}}$=1.82$\pm$0.01~$\mu_{\text{B}}$/Cu$^{2+}$ where $C_{\text{1D}}$=$g^2\left(\text{N}_{\text{A}}\mu_{\text{B}}^2/4\text{k}_{\text{B}}\right)$, agrees with the value of the free S=1/2 Cu$^{2+}$ ion.

\subsubsection{Heat Capacity}\label{sec:HeatCapacity}

\begin{figure}
\centering
\hspace*{-0.2cm} 
\begin{tabular}{c @{\qquad} c }
\includegraphics[width=0.83\columnwidth,keepaspectratio]{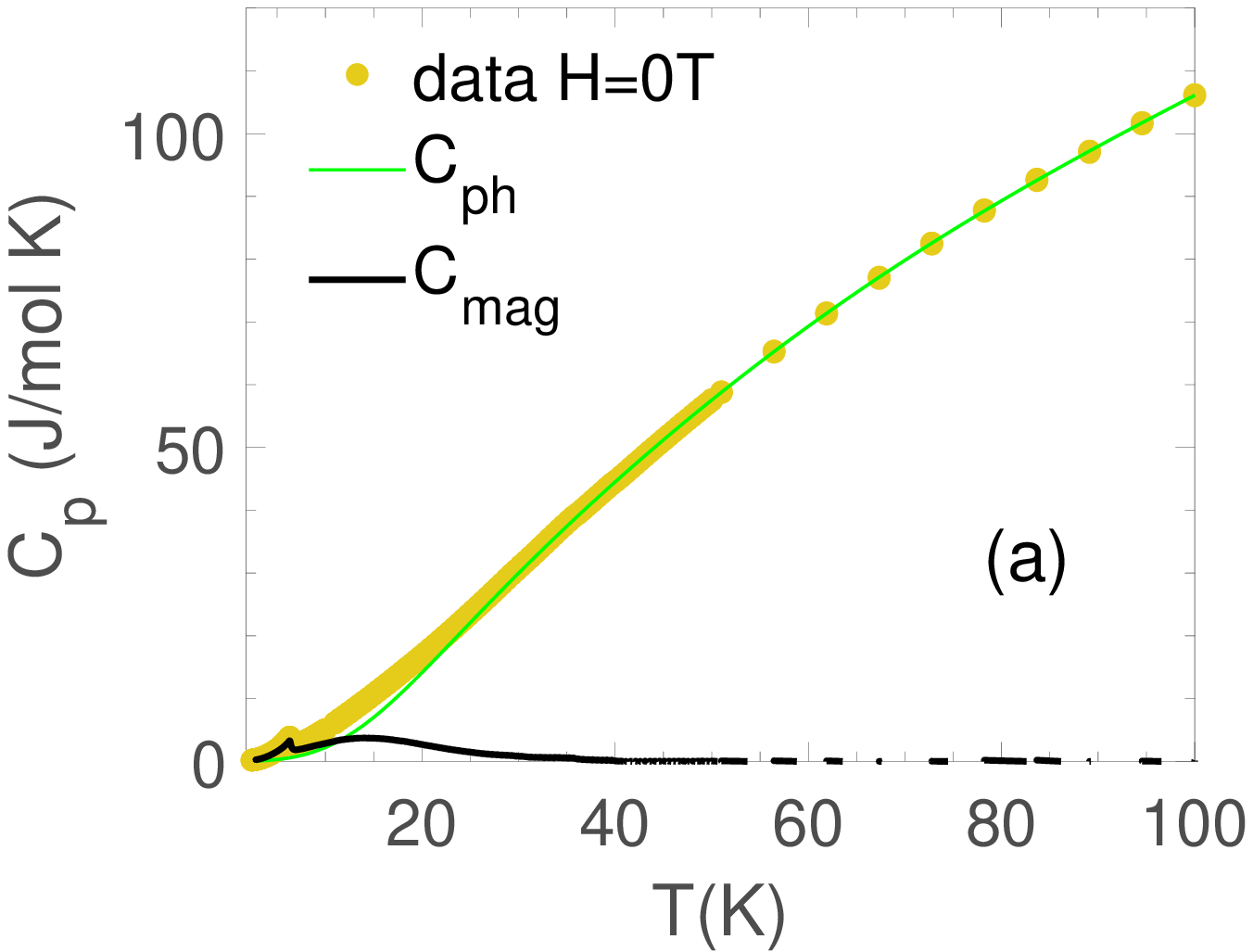} \\

\hspace*{+0.45cm} 
\includegraphics[width=0.93\columnwidth,,keepaspectratio]{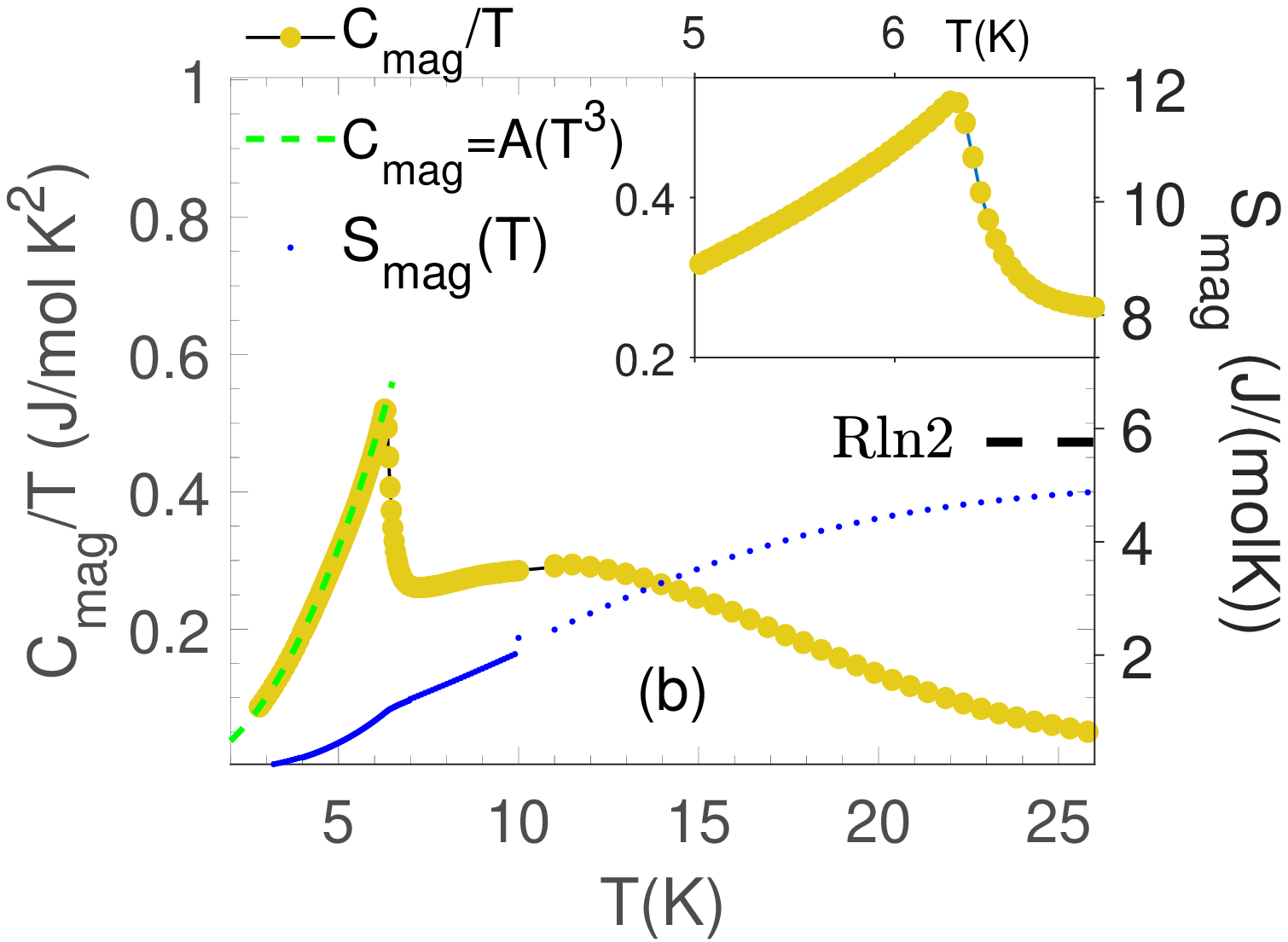}  \\
 
 \end{tabular}
\caption{(a) Heat capacity data collected as a function of temperature from a powder sample of BaCuTe$_2$O$_6$ in zero applied magnetic field. The data were fitted with a linear combination of 3 Debye terms above $T=$40~K, to model the lattice heat capacity (sold green line). The  phononic parameters obtained from fitting Eq.~\ref{Debye model_bcto} to the data are: $c_{1}$=2.2(2) with $T_{Di}$=128(5), $c_{2}$=2.96(6) with $T_{Di}$=333(15) and $c_{3}$=37(3) with $T_{Di}$=1905(2). The magnetic heat capacity, $C_{\text{mag}}$, was obtained by subtracting the phononic contribution from the data (solid black line). (b) The low temperature $C_{\text{mag}}/T$, is given by the yellow symbols (left-hand y-axis), as a function of temperature and reveals a $\lambda$-anomaly at $T_{\text{N}}$=6.32~K. The green dashed line is a fit of $C_{\text{mag}}$ below $T_N$, by the power law $C_{\text{mag}}$=$A(T^3)$. The temperature-dependence of the magnetic entropy is given by the black dots (right-hand $y$-axis).}
\label{bcto_Cp_fitMag}
\end{figure}

The heat capacity $C_\text{P}$, of BaCuTe$_2$O$_6$ was investigated to provide further information about the magnetic properties. Data were collected on a powder sample in zero applied magnetic field, over the temperature range 2 to 100~K. Figure~\ref{bcto_Cp_fitMag}(a) presents the results where a sharp $\lambda$-anomaly appears at $T\approx6$~K suggesting a 2$^{nd}$ order phase transition in agreement with the peak observed in the dc-susceptibility measurement. At higher temperatures the heat capacity increases with temperature. The $C_{\text{P}}$ of BaCuTe$_2$O$_6$ has two contributions, phononic and magnetic, the magnetic contribution is expected at temperatures below $T_{\text{CW}}\approx36$~K while the phonons dominate at higher temperatures. To model the lattice contribution a linear combination of $n$ Debye integrals was fitted to the data for $T>40$~K \cite{Kittel1996} using the expression  
\begin{equation}
\begin{aligned}\label{Debye model_bcto}
 C_{\text{ph}}(T)=9~R~\sum_{i=1}^{n}c_i\left(\frac{T}{T_{\text{D}i}}\right)^3\int_{0}^{T_{\text{D}i}/T} ~\frac{x^4e^x}{(e^x-1)^2}~dx, 
\end{aligned}
\end{equation}
where $R$ is the universal gas constant and $c_i$ is the $i^{th}$ coefficient with Debye temperature $T_{\text{D}i}$. The best fit was achieved for $n$=3 as shown by the green line in Fig.~\ref{bcto_Cp_fitMag}(a) and the fitted parameters are given in the figure caption. 

The pure magnetic $C_{\text{mag}}$ contribution is shown by the black line and was obtained by subtracting the $C_{\text{ph}}$ part from the data. Fig.~\ref{bcto_Cp_fitMag}(b) presents $C_{\text{mag}}/T$ as a function of $T$ (yellow points, left-hand y-axis) focusing on the low temperature region where the broad maximum at $T_{\text{m}}$=10.63(4)~K is associated with short-range spin correlations while the sharp $\lambda$-anomaly at $T_{\text{N}}$=6.32~K indicates the transition to long-range magnetic order. The inset shows the region around $T_{\text{N}}$ in detail, and reveals a single transition in contrast to SrCuTe$_2$O$_6$ which has two distinct transitions. In order to estimate the exchange interaction J3 from the heat capacity data, we use the relation for a spin-1/2, Heisenberg antiferromagnetic chain \cite{Johnston2000}, J3$/k_\text{B} = T_{\text{m}}/0.3071$, which yields J3$/k_{\text{B}}$=34.6(1)~K confirming the result from the $\chi_{\text{dc}}$ measurement. Finally, the heat capacity below the transition follows a $T^{3}$ power law (shown by the green dashed line), which suggests the presence of a three-dimensional spin-wave dispersion at lowest energies and temperatures. 

The magnetic entropy $S_{\text{mag}}$, was calculated from the heat capacity and is plotted as a function of temperature in Fig.~\ref{bcto_Cp_fitMag}(b) (blue points, right-hand y-axis). It increases with temperature reaching a value of 4.87~Jmol$^{-1}K^{-1}$ at $\sim$25~K above which it starts to approach the maximum value of $S_{\text{mag}}$=$R\text{ln}2$=5.76~Jmol$^{-1}K^{-1}$ expected for a spin-1/2 magnet. It should be noted that over the temperature range from 2~K to just above the transition at 7~K, only 21.2$\%$ of the full entropy is recovered, this is a feature typical of low-dimensional and / or frustrated magnets.                                  

\subsection{Magnetic structure}

\begin{figure}
\centering
\begin{tabular}{c @{\qquad} c }
\hspace*{0.2cm}
\vspace*{0.5cm}
\includegraphics[width=0.83\linewidth,keepaspectratio]{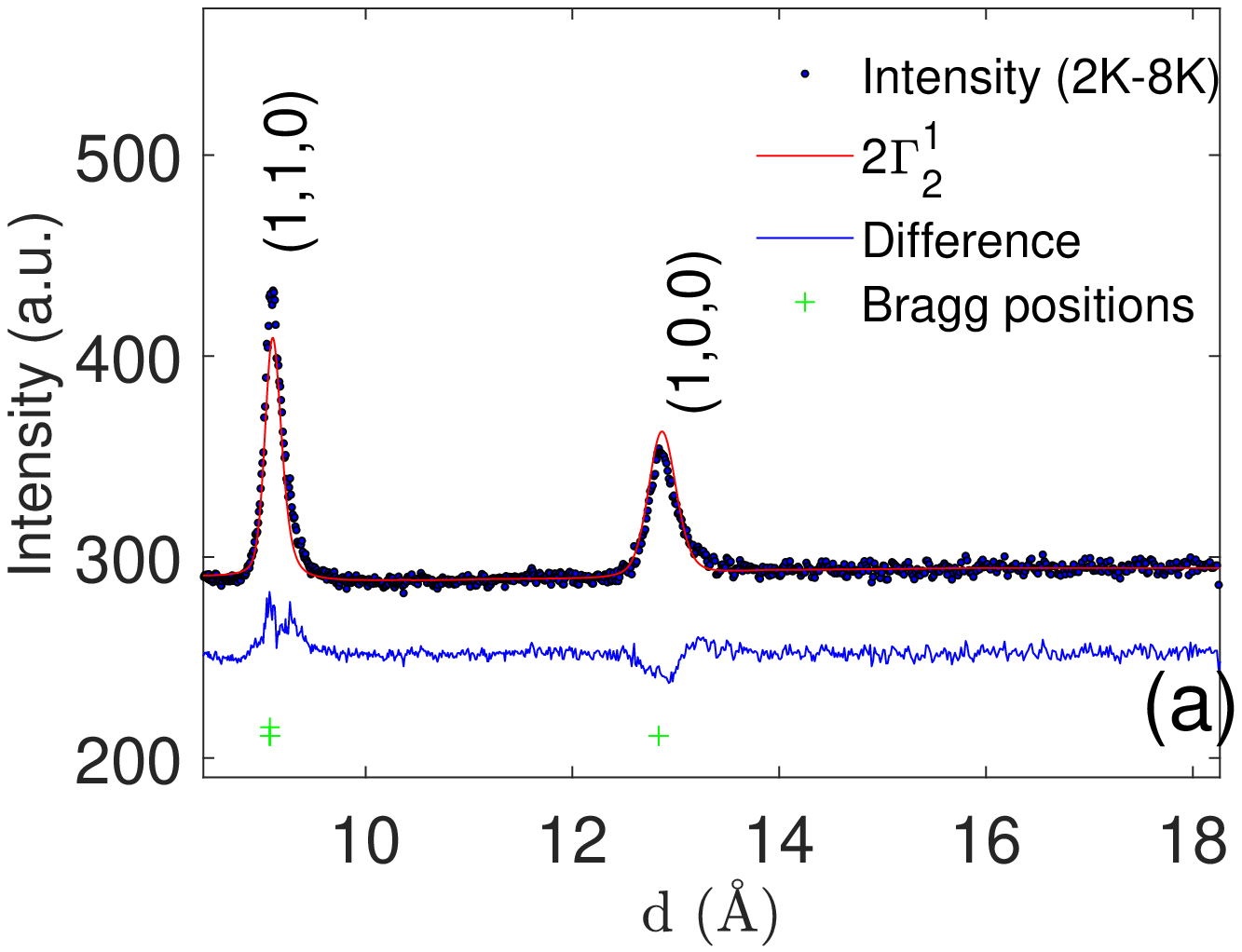}  \\
\hspace*{0.2cm}
\includegraphics[width=0.83\linewidth,keepaspectratio]{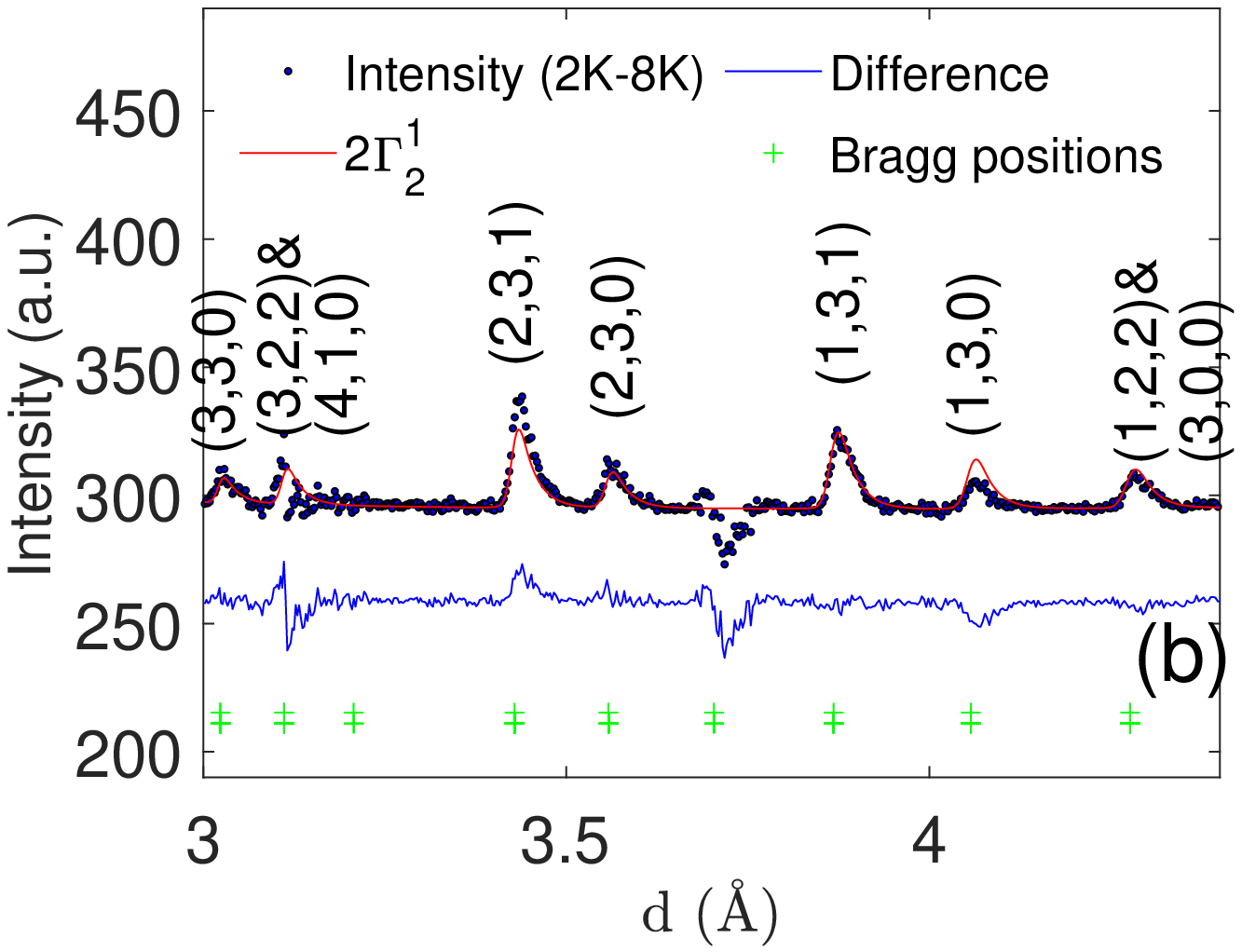}  \\
 \end{tabular}  
\caption{Rietveld refinement of the magnetic diffraction pattern collected on (a) detector banks 1 and 10 showing the high $d$-spacing peaks, and (b) detector banks 2 and 9 showing the low $d$-spacing peaks, of the WISH diffractometer for a powder sample of BaCuTe$_2$O$_6$. The black circles give the result of subtracting the 8~K dataset from the 2~K dataset, the red line gives the refined fit of the data assuming the IR $2\times \Gamma_2^1$ ($\chi^2$~=~3.1) while the blue line gives the difference between theory and experiment.}
\label{bcto_wish_Fit}
\end{figure}

We now investigate the nature of the magnetic order occurring below $T_{\text{N}}$ in BaCuTe$_2$O$_6$. The neutron diffraction data for the magnetic structure determination were collected on the WISH diffractometer which is an effective instrument for magnets with small ordered moments \cite{Wish}. The difference pattern (the dataset at 2~K after subtraction of the 8~K dataset) is presented in Fig.~\ref{bcto_wish_Fit} over selected $d$-space regions and a list of the peaks that appear after this subtraction is presented in Table~\ref{magnetic Bragg peaks} (the difference patterns for the other detector banks are given in Appendix~\ref{sec:MagPeaks}). The Bragg peak at $d$=12.83~$\AA$ corresponds to the forbidden nuclear reflection (1,0,0), and is a clear indication of the magnetically ordered state at 2~K. All peaks can be indexed by the magnetic ordering vector, $k$=(0,0,0), suggesting a commensurate magnetic structure.

\begin{figure}
\centering
\begin{tabular}{c @{\qquad} c }
\hspace*{0.2cm} 
\includegraphics[width=0.83\linewidth,keepaspectratio]{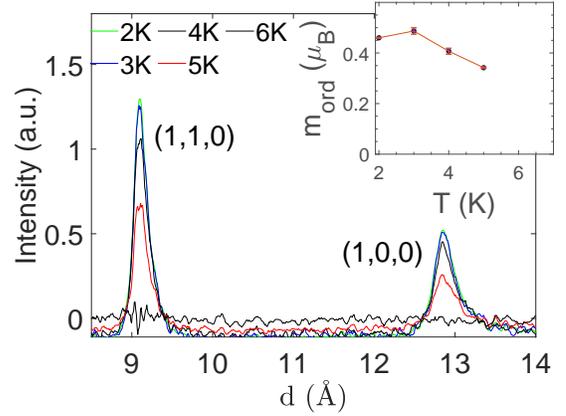}  \\
 \end{tabular}  
\caption{Temperature dependence of the magnetic Bragg peaks obtained by subtracting the neutron diffraction pattern measured at 8~K from the $T$=2, 3, 4, 5 and 6~K neutron diffraction patterns. The datasets were collected from detector banks 1 and 10 of WISH. In the inset the fitted ordered moment m$_{\text{ord}}$ is presented as a function of temperature. The values were found after refining the data with the $2\times \Gamma_2^1$ IR.}
\label{bcto_wish_MagPeaks_AllTemp}
\end{figure}

Figure \ref{bcto_wish_MagPeaks_AllTemp} presents the (1,0,0) and (1,1,0) ($d_{(110)}=9.07\AA$) magnetic reflections at several different temperatures from 2~K to 6~K (after subtraction of the 8~K dataset). The intensity decreases with increasing temperature and disappears at 6~K suggesting that BaCuTe$_2$O$_6$ enters a long range magnetically ordered state below 6~K in agreement with the heat capacity and susceptibility data. Within the temperature steps that were measured no indication of a second transition is found. 

\begin{table}\label{magnetic Bragg peaks}
\caption{\label{magnetic Bragg peaks} Peaks that appear after the subtraction of the $T$=8~K from the $T$=2~K datasets measured on the WISH detector banks 1, 2, 9 and 10.}
\begin{center}
\resizebox{\columnwidth}{!}{%
\begin{tabular}{ |c|c|c|c|c|c|c|}
\hline
(h, k, l) &(1,0,0) &(1,1,0) &(1,1,1) &(0,1,2) &(1,1,2) &(1,2,2) or (3,0,0)  \\
\hline
d($\AA$) &12.8328  &9.07409 &7.40896 &5.73896  &5.23784  &4.27757 \\
\hline
 \multicolumn{7}{|c|}{ } \\
\hline
(h, k, l) &(0,1,3) &(1,1,3) &(0,2,3) &(1,2,3)&(0,3,3)&(4,1,0) or (3,2,2)\\
\hline
d($\AA$) &4.05806 & 3.86920&3.55915 &3.42968  &3.02470 & 3.11099\\
\hline
\end{tabular}
}
\end{center}
\end{table}  

We employed the standard representation theory approach \cite{Yu2011}, assuming that the magnetic structure is transformed by a single irreducible representation of the paramagnetic space group P$4_1$32. The corresponding basis vectors were calculated using the software BASIREPS within the FullProf Suite \cite{Fullprof1993}. The magnetic representation of the Cu$^{2+}$ ion (Wyckoff site 12$d$) can be reduced into the following five irreducible representations (IRs) of the P$4_1$32 space group, associated with the $\boldmath{k}$=(0,0,0) propagation vector:
\begin{equation}
\Gamma_\text{mag\ ,Cu} = 1 \times \Gamma_1^1 + 2\times \Gamma_2^1 + 3\times \Gamma_3^2 + 4\times  \Gamma_4^3+ 5\times \Gamma_5^3
\label{eq:IRs}
\end{equation}
Each of the terms corresponds to one of the irreducible representations $(t\times \Gamma_o^d)$ where the coefficient $t$ gives the number of times it occurs, the subscript $o$ gives the order of the representation and the superscript $d$ gives the dimensionality. 
The possible magnetic ground states are obtained from the combination of the basis vectors of each of the IRs, and the corresponding magnetic space groups were evaluated using the ISODISTORT software \cite{Campbell2006}. 

The candidate magnetic structures, whose magnetic space groups are consistent with the observed reflection conditions and are maximal isotropy subgroups of P$4_1$32, were selected and systematically tested in the refinement procedure. As discussed in Appendix~\ref{sec:MagPeaks}, the (H,H,0) reflections (where H is an odd integer) are always present for BaCuTe$_{2}$O$_{6}$ while the (H,H,H) peaks are not observed except for (1,1,1). This matches best with the cubic magnetic structure associated with the one-dimensional IR $2\times \Gamma_2^1$  having the P$4_1$'32' symmetry, for which (H,H,0) are allowed and (H,H,H) are forbidden. We believe the presence of the (1,1,1) reflection is an artifact arising from the subtraction of the very strong nuclear contribution. This conclusion is supported by the absence of any magnetic contribution at the substantially weaker nuclear reflections (2,2,2) and (3,3,3) (see Appendix~\ref{sec:MagPeaks}). 

The magnetic structure of BaCuTe$_2$O$_6$ was therefore refined using $2\times \Gamma_2^1$ where the crystallographic parameters were fixed to the refined values found from the neutron diffraction measurement performed at 8~K. Fig.~\ref{bcto_wish_Fit} shows the best fit (red line) to the neutron diffraction data collected at 2~K (black circles) giving $\chi^2$~=~3.1. The difference between the fit and data is shown by the blue line and shows good agreement except for the (1,1,1) peak. The ordered moment at 2~K was found to be $m_\text{ord}$=0.46~$\mu_\text{B}$/Cu$^{2+}$. It is significantly suppressed compared to the value for the free Cu$^{2+}$ ion of 1~$\mu_\text{B}$. The suppression of the ordered moment is attributed to the presence of low-dimensional and/or frustrated antiferromagnetic interactions which give rise to strong spin fluctuations. 

The $2\times \Gamma_2^1$ IR found for BaCuTe$_2$O$_6$ contrasts with the $1\times \Gamma_1^1$ IR, recently found for the isostructural compound SrCuTe$_2$O$_6$ \cite{Chillal2020,Saeaun2020}, where the latter IR forbids both the (H,H,0) and (H,H,H) reflections. We therefore also refined the diffraction pattern of BaCuTe$_2$O$_6$ using the $1\times \Gamma_1^1$ IR which gave the much worse goodness-of-fit of $\chi^2$~=~8.5. As shown in Appendix~\ref{sec:Gamma1}, $1\times \Gamma_1^1$ fails to reproduce the intensity at the (1,1,0) peak and generally provides a much poorer fit over all $d$-spacing ranges than $2\times \Gamma_2^1$. Hence, contrary to the findings for SrCuTe$_2$O$_6$, $2\times \Gamma_2^1$ is the correct IR for BaCuTe$_2$O$_6$. The difference in the magnetic structure of these two compounds may be attributed to the difference in the ionic sizes of Ba$^{2+}$ and Sr$^{2+}$, with the ionic radius of Ba$^{2+}$ being 14\% larger than that of Sr$^{2+}$, which could alter the bond distances and angles and hence the strength of the magnetic interactions (see Appendix~\ref{sec:StructDet}).

We now discuss the higher order IRs, $3\times \Gamma_3^2$, $4\times  \Gamma_4^3$ and $5\times \Gamma_5^3$. The maximal magnetic subgroups associated with these IRs are inconsistent with the magnetic reflections of BaCuTe$_2$O$_6$. $3\times \Gamma_3^2$  and $4\times  \Gamma_4^3$ have a systematic absence of the (2n,0,0), (0,2n+1,0) and (0,0,2n+1) magnetic peaks (where n is an integer), thus we would expect to see (2,0,0) in our powder measurement. Similarly $5\times \Gamma_5^3$ is not compatible since it predicts the systematic absence of (H,H,0) peaks whereas both (1,1,0) and (3,3,0) are present. We did not consider the candidate magnetic structures whose magnetic space groups are not maximal isotropy subgroups of P4$_1$32. These magnetic structures require strongly first order phase transitions and therefore are unlikely to occur in BaCuTe$_2$O$_6$.

\begin{figure}
\centering
\hspace*{-1.0cm}
\includegraphics[width=0.9\linewidth,keepaspectratio]{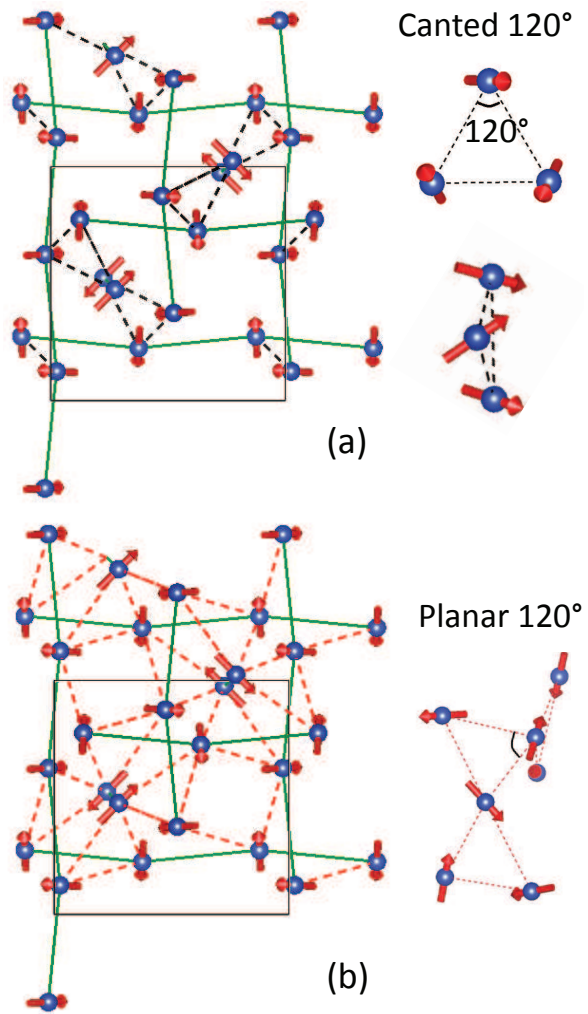}
\caption{Visualization of the magnetic structure of  BaCuTe$_2$O$_6$ found by fitting the diffraction patterns measured at 2~K by the $2\times \Gamma_2^1$ IR. Only the Cu$^{2+}$ are shown represented by the blue spheres, while the red arrows indicate the directions of the ordered spin moments. (a) includes the first J1 (dotted black lines), and third J3 (green lines), nearest neighbor interactions while (b) includes the second J2 (dotted red line), and third J3 (green lines) nearest neighbor interactions. To the right of panel (a) [(b)], the triangles formed by J1 [J2] are presented. This figure was produced using the VESTA software \cite{VESTA}.}
\label{bcto_ord_moment_visual}
\end{figure}

The magnetic structure of BaCuTe$_2$O$_6$ is depicted in Fig.~\ref{bcto_ord_moment_visual} where only the Cu$^{2+}$ ions are shown (blue spheres) and the red arrows indicate the directions of their ordered spin moments. The magnetic interactions have also been included and by inspecting the relative alignment of the spins connected by these bonds, information about the size and sign of these interactions can be deduced. Figure~\ref{bcto_ord_moment_visual}(a) includes the first J1, and third J3, nearest neighbor interactions, represented by the dotted black and solid green lines respectively. It is immediately apparent that the spins of the Cu$^{2+}$ ions connected by J3 form antiferromagnetic chains, clearly identifying this interaction as antiferromagnetic. An isolated triangle formed by J1 is also presented separately on the right in planar and side views. The spins are non-collinear and have a strongly canted structure. Their components in the plane of the triangle have a 120$^{\circ}$ arrangement typical of an antiferromagnetic interaction, while in contrast their components perpendicular to the plane are ferromagnetically aligned. This suggests that J1 is too weak to influence the magnetic structure and it is unclear whether it is ferromagnetic or antiferromagnetic. Figure~\ref{bcto_ord_moment_visual}(b) includes the second J2 (dotted red lines), and third J3 (solid green lines), nearest neighbor interactions. Three of the corner-sharing triangles formed by J2 are shown on the right. The spins form a 120$^{\circ}$ arrangement in the plane of every triangle with zero out-of-plane component showing that this interaction is antiferromagnetic and sufficiently strong to influence the magnetic order. Altogether the magnetic structure of BaCuTe$_2$O$_6$ reveals that J2 and J3 are antiferromagnetic while J1 is very weak.

Assuming spin chains created by J3 which are coupled together by J2, a rough estimate of the size of J2 can be found from the ordered moment $m_{ord}$ via the expression $m_{ord}=1.0197\sqrt(J2/J3)$ \cite{schulz}. Using J3=34~K, we get J2=6.91~K. Since this expression applies to unfrustrated interchain coupling, we anticipate that it provides only an approximate lower limit on the frustrated J2 interchain coupling of BaCuTe$_{2}$O$_{6}$.

\subsection{Magnetic excitations}

\begin{figure}
\centering
\begin{tabular}{c @{\qquad} c }
\includegraphics[width=0.80\linewidth,keepaspectratio]{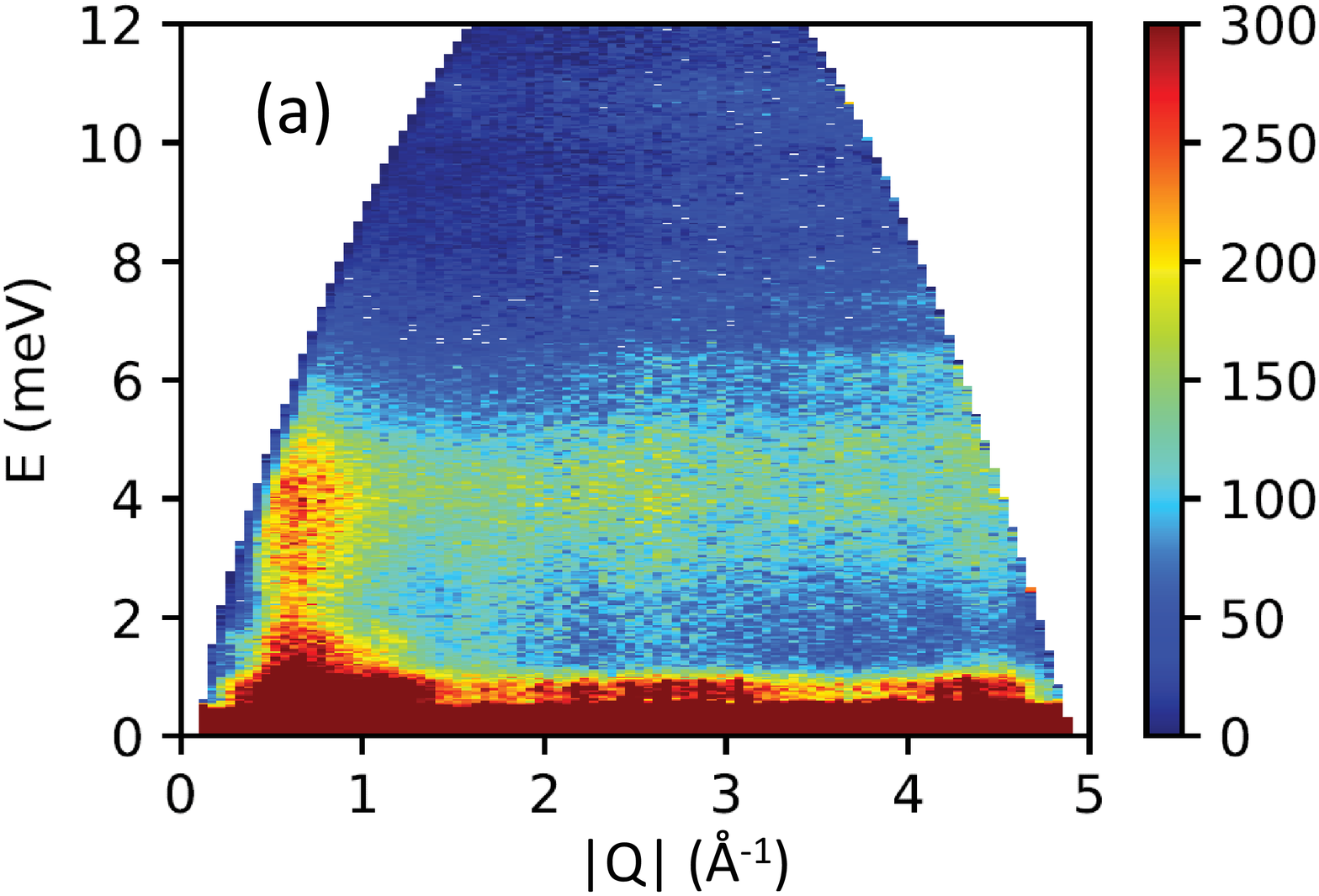}  \\
\includegraphics[width=0.80\linewidth,keepaspectratio]{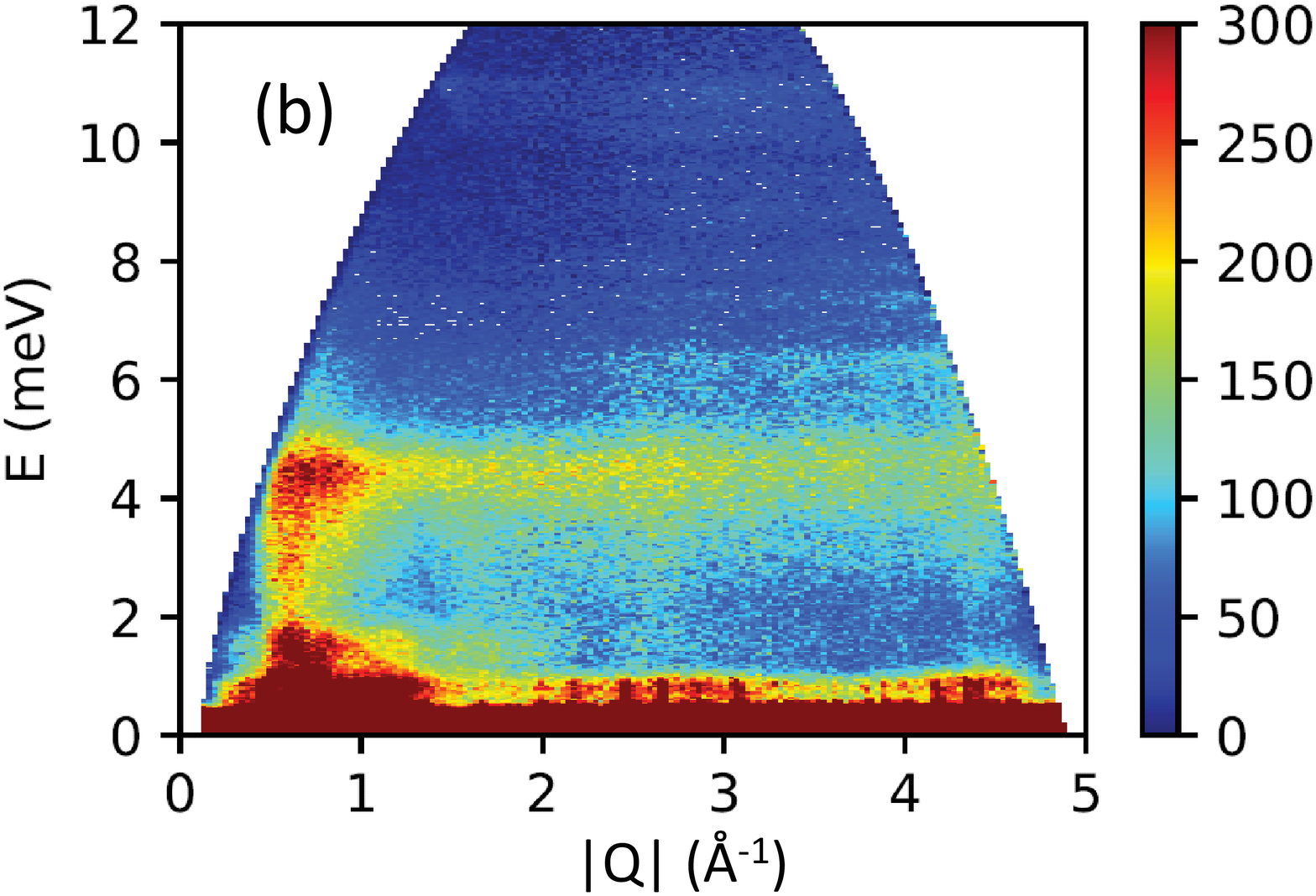}    \\
\hspace*{-1.0cm}
\includegraphics[width=0.68\linewidth,keepaspectratio]{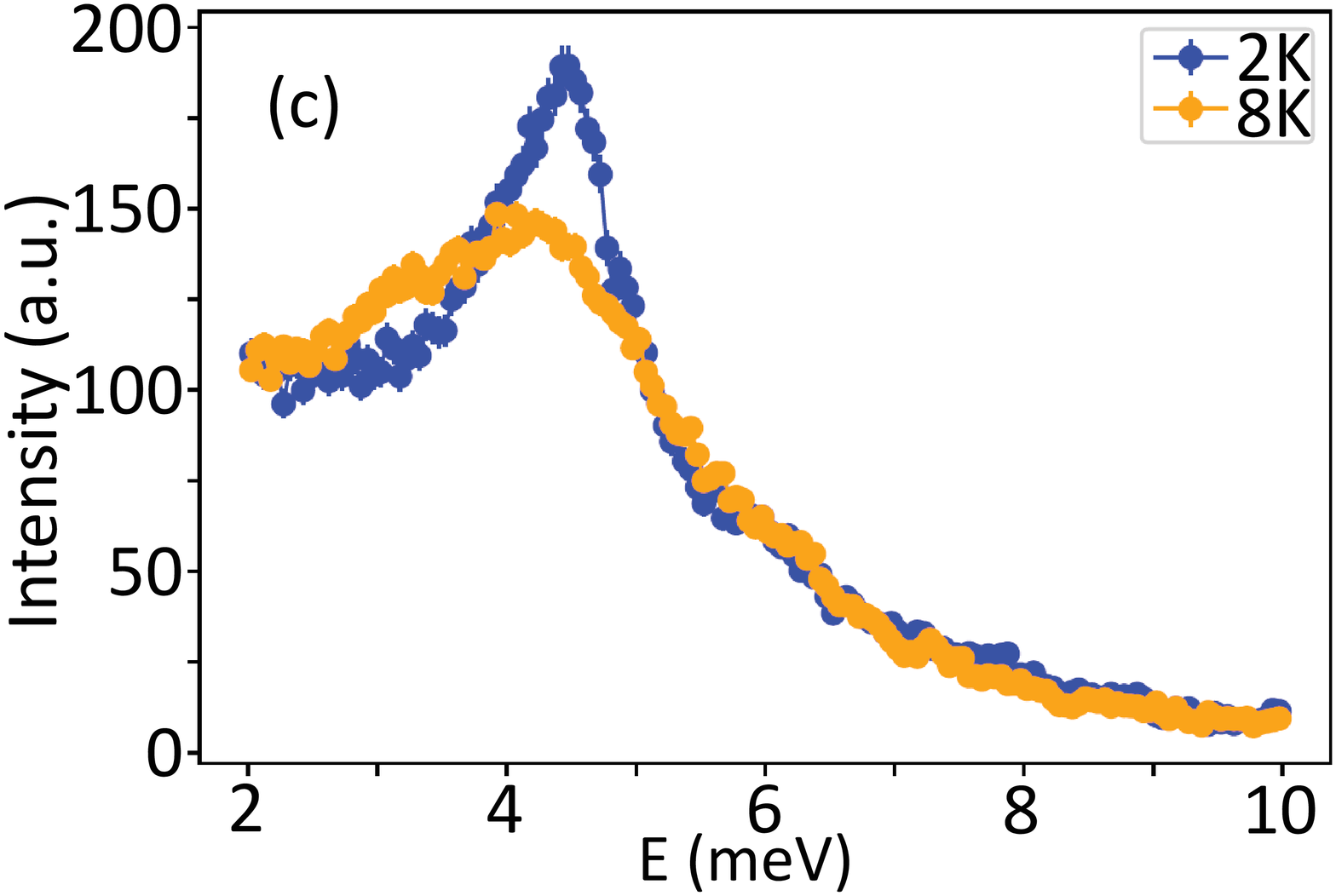} \\
\end{tabular}
\caption{Inelastic neutron scattering spectra as a function of energy and wavevector transfer. The data were measured using the LET ToF spectrometer on a powder sample at $T$=8~K (a) and $T$=1.8~K (b) with an incident energy, $E_{\text{i}}$=14.10~meV. No background has been subtracted from the data. (c) Energy cut through the data integrated over the wavevector range 1.0~$\AA^{-1}$$<$$|Q|$$<$2.0~$\AA^{-1}$ and plotted as function of energy. A broad maximum at $\approx$4~meV at $T$=8~K (yellow symbols) becomes sharper at 4.5~meV in the ordered state at $T$=1.8~K (blue symbols). With increasing energy, the signal drops gradually and seems unaffected by the temperature.}
\label{LET_powder_EQ}
\end{figure}

\begin{figure}
\centering
\begin{tabular}{c @{\qquad} c }
\hspace*{-0.5cm} 
\includegraphics[width=1.1\linewidth,keepaspectratio]{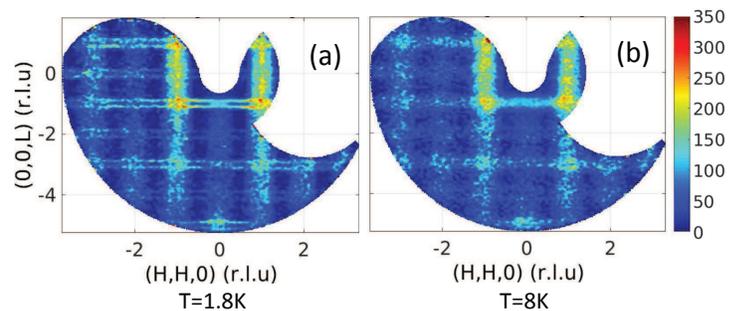}
\end{tabular}
\caption{ Inelastic neutron scattering plotted as a function of wavevector transfer in the [H,H,L] plane at a constant energy transfer of 2.0~meV. The data was measured using the LET ToF spectrometer on a single crystal sample at $T$=1.8~K (a) and $T$=8~K (b) using an incident energy of $E_{\text{i}}$=4.81~meV. The data was integrated over the energy transfer range 1.8~meV$<$$E$$<$2.2~meV and over wavevector in the vertical out of plane direction by [$\pm$0.2,$\mp$0.2,0]. No background subtraction has been performed and only a subtle smoothing is implemented.}
\label{fig:LET_sc_QQ_collection}
\end{figure}

We now investigate the magnetic excitations of BaCuTe$_2$O$_6$. Fig.~\ref{LET_powder_EQ} presents inelastic neutron scattering (INS) measured on the LET time-of-flight (ToF) spectrometer for a powder sample above the N\'{e}el temperature at 8~K (Fig.~\ref{LET_powder_EQ}(a)) and below $T_N$ at 1.8~K (Fig.~\ref{LET_powder_EQ}(b)). At 8~K, distinctive diffuse scattering occurs for energies below $\approx7$~meV which is identified as magnetic because its intensity decreases with increasing wavevector transfer $|Q|$ in accordance with the magnetic form factor. There are other much weaker bands at higher energies e.g. at 9~meV$<$$E$$<$11~meV, which are identified as having phononic origin since their intensity increases with increasing $|Q|$. The magnetic signal is generally sharper below $T_{\text{N}}$ at 1.8~K and additional signal appears below $E$=1.5~meV. Fig.~\ref{LET_powder_EQ}(c) presents a cut through the data along energy at 1.8~K and 8~K with wavevector integrated over the range 1.0~$\AA^{-1}$$<$$|Q|$$<$2.0~$\AA^{-1}$ which shows a clear sharpening and intensity increase of the excitations due to the ordering of the magnetic moments below $T_N$.
 
\begin{figure}
\centering
\begin{tabular}{c @{\qquad} c }
\includegraphics[width=0.98\linewidth,keepaspectratio]{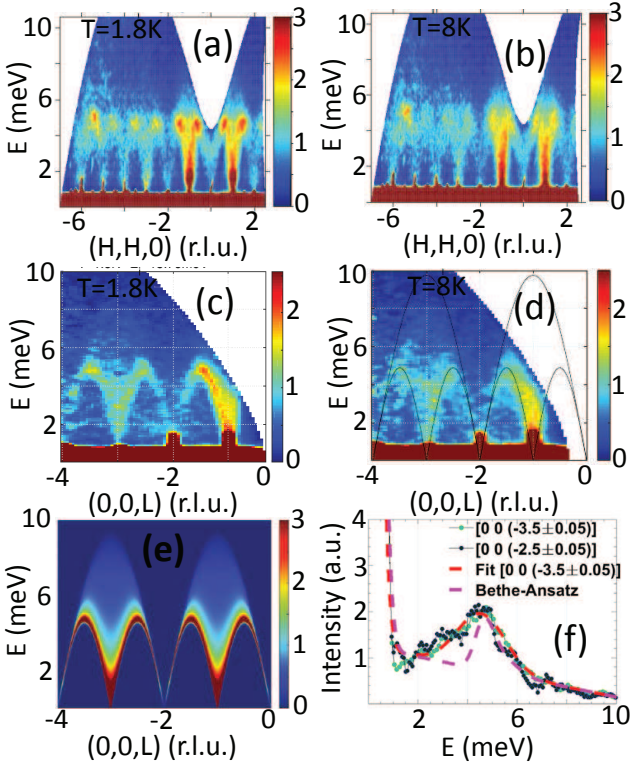}
\end{tabular}
\caption{ (a-d) Inelastic neutron scattering measured on a single crystal sample plotted as a function of energy and wavevector transfer without  background subtraction. The data were collected on the LET ToF spectrometer using an incident neutron energy of $E_{\text{i}}$=13.75~meV (energy resolution, 0.78~meV). (a) and (b) show the spectra along [H,H,0] collected at $T$=1.8~K and $T$=8~K respectively. The data have been integrated by $\pm$0.2~r.l.u. over the [0,0,L] direction and by $\pm$0.5~r.l.u. over the [K,-K,0] vertical out-of-plane direction. (c) and (d) show the spectra along [0,0,L] collected at $T$= 1.8~K and $T$=8~K respectively. The data have been integrated by $\pm$0.2~r.l.u. over the [H,H,0] and [K,-K,0] directions. The ranges of integration were selected to achieve an adequate resolution and intensity. (e) Resolution corrected dynamical structure factor calculated via the algebraic Bethe-Ansatz for J3=2.90(6)~meV. (f) Cuts along energy through the data collected with $E_\text{i}$=13.75~meV and $T$=1.8~K at L=-2.5 and -3.5 with integration over L of $\pm$0.05, and over [H,H,0] and [K,-K,0] of $\pm$0.2~r.l.u. The red line gives the fit of a Gaussian to the L=-3.5 cut in order to extract the maximum of the lower boundary of the spinon continuum (red line). The magenta line gives the related cut through the theoretical dynamical structure factor calculated via the algebraic Bethe-Ansatz for J3=2.90(6)~meV at L=-3.5. The fits include a sloping background and the elastic incoherent scattering peak.}
\label{fig:LET_sc_QE_collection}
\end{figure}

In order to learn more about the spin dynamics of  BaCuTe$_2$O$_6$, further INS data were collected on the LET spectrometer from a single crystal sample. Fig.~\ref{fig:LET_sc_QQ_collection} presents the spectrum collected for an incident energy of $E_{\text{i}}$=4.81~meV, integrated over the narrow energy transfer range 1.8~meV$<$$E$$<$2.2~meV, and plotted in the [H,H,L] reciprocal plane. The data in Fig.~\ref{fig:LET_sc_QQ_collection}(a) were measured at $T$=1.8~K and in Fig.~\ref{fig:LET_sc_QQ_collection}(b) at $T$=8~K. At 8~K a highly symmetric pattern is revealed with two sets of parallel diffuse scattering streaks which are perpendicular to each other. Streaks of scattering observed in constant-energy slices indicate dispersionless excitations along the streak direction and are typical features of low-dimensional magnetism where the excitations are dispersionless in certain crystallographic directions. In the case of a magnet with one-dimensional interactions, the spectrum varies only along the chain direction and is dispersionless perpendicular to the chain. 

In BaCuTe$_2$O$_6$ the only interaction which could give rise to the observed features is J3 which couples the Cu$^{2+}$ ions antiferromagnetically into mutually perpendicular chains along the a, b and c directions. In contrast J1 would produce a zero dimensional magnetism while J2 would result in three-dimensional magnetism. The first set of streaks which appears at integer values of L and run parallel to the [H,H,0] direction, arise from the spin-chains along the c-axis. The second set, which occurs at integer values of H and are parallel to [0,0,L], originate from the spin chains alonh the a- and b-axes. It is a characteristic feature that at odd values of H and L the streaks are systematically weaker than at even values. This scattering pattern sharpens below the transition at 1.8~K (Fig.~\ref{fig:LET_sc_QQ_collection}(a)) where the streaks split into closely spaced double-streaks.

Energy slices as a function of wavevector parallel to the [H,H,0] and [0,0,L] directions can provide further insight into the spin dynamics. Figures~\ref{fig:LET_sc_QE_collection}(a-d) presents ToF INS spectra measured on LET with incident energy $E_{\text{i}}$=13.75~meV, at $T$=1.8~K (Fig.~\ref{fig:LET_sc_QE_collection}(a,c)) and $T$=8~K (Fig.~\ref{fig:LET_sc_QE_collection}(b,d)). The constant energy cuts along [H,H,0] (Fig.~\ref{fig:LET_sc_QE_collection}(a,b)) show dispersive features which are associated with the chains parallel to the a- and b-axes. Gapless, V-shaped excitations disperse up from the odd integer L and H reciprocal lattice points, while they are considerably weaker at even integer L and H values. The excitations are not sharp but form a continuum which broadens over wavevector with increasing energy transfer achieving strong intensity at $E\approx$5~meV above which they weaken and then disappear for energies greater than 9~meV. The spectrum appears to become more intense and better defined below the N\'{e}el temperature than above.  Figures~\ref{fig:LET_sc_QE_collection}(c, d) show the continuum along the [0,0,L] direction collected at 1.8~K (Fig.~\ref{fig:LET_sc_QE_collection}(c)) and at 8~K (Fig.~\ref{fig:LET_sc_QE_collection}(d)). These features are associated with the chains parallel to the c-axis and disperse up from the odd integer L reciprocal lattice points but are significantly weaker at even L.

The high energy excitations of BaCuTe$_2$O$_6$ resemble the spectrum of a spin-1/2 Heisenberg antiferromagnetic chain where they take the form of a broad multi-spinon continuum rather than conventional spin-waves which would form sharp dispersive modes. Fig.~\ref{fig:LET_sc_QE_collection}(e) shows the theoretical dynamical structure factor for the multi-spinon continuum according to the Bethe-Ansatz \cite{Muller1981, Caux2005} which can be compared to the data. Note that because the structural unit cell contains two Cu$^{2+}$ ions per chain, the two-spinon continuum emerges from odd integer reciprocal lattice points rather than at half-integer values. The spinon continuum is contained between a lower and upper boundary given by the relations $E_L$=($\pi$/2)J3$|\sin$($\pi L)|$ and $E_U$=$\pi$ J3$|\sin$($\pi L/2)|$ respectively. The intensity is  strongest at the lower boundary, and for half-integer values of L, a peak would be observed in a cut along energy marking the onset of the continuum at $E_L(max)$=$\pi$J3/2, this can thus be used to extract the value of the intrachain interaction J3.

Figure \ref{fig:LET_sc_QE_collection}(f) shows cuts through the 1.8~K dataset as a function of energy at L=2.5 and 3.5 r.l.u. with a narrow integration range over L of $\pm0.05$~r.l.u. The data reveal a peak at $ E_L(max)$=4.56(9)~meV, as found by fitting a Gaussian on a sloping background to the data (red line). This gives the size of the antiferromagnetic intrachain interaction as J3=2.90(6)~meV or 33.7(7)~K. The black line in Fig.~\ref{fig:LET_sc_QE_collection}(d) gives the boundaries of the two-spinon continuum calculated using the Bethe-Ansatz for J3=2.90(6)~meV. The calculations are in good agreement with our data. Moreover, the J3 value is found to be similar to the value extracted from dc-susceptibility of 34.40(4)~K and from heat capacity of 34.6(1)~K.  The consistency between the size of J3 from different experimental measurements and the qualitative agreement of the experimental INS data with the Bethe-Ansatz theory, including the observation of continuum scattering, provide proof that the J3 interaction is the dominant interaction and that BaCuTe$_2$O$_6$ is a good realization of the one-dimensional spin-1/2 Heisenberg antiferromagnet where the chains run parallel to the a-, b-, and c-axes. 

Finally, although J1 and J2 are clearly much weaker that J3, one or both of them must still be finite. As shown in Fig.~\ref{bcto_ord_moment_visual}, these interactions connect the chains together. In the absence of interchain coupling, a one-dimension magnet cannot develop long-range magnetic order according to the Mermin-Wagner theorem \cite{Mermin}. Since BaCuTe$_2$O$_6$ orders antiferromagnetically below $T_\text{N}$=6.3~K the presence of J1 and/or J2 is confirmed.

\section{Summary and Conclusions}

In summary we have performed a detailed investigation of the new quantum magnet BaCuTe$_2$O$_6$. We succeeded to make the first powders using the solid state reaction method and single crystals using the optical floating zone technique, and our characterizations suggest that our samples are pure and of high quality. Our synchrotron X-ray powder diffraction measurements reveal that BaCuTe$_2$O$_6$ crystallizes in the cubic space group P4$_1$32 (a=12.83 $\AA$) and is isostructural to the other members of this family, SrCuTe$_2$O$_6$ and PbCuTe$_2$O$_6$. Neutron powder diffraction performed using a cryostat, shows that this structure is stable down to $T$~=~0.5~K. DC susceptibility measurements find a negative Curie-Weiss temperature $T_\text{CW}$~=~-32.8(1)~K suggesting predominantly antiferromagnetic spin interactions, while a transition to long-range antiferromagnetic order occurs at $T_{\text{N}}$~=~6.31~K. The much lower value of $T_{\text{N}}$ compared to $T_\text{CW}$ as well as the broad maximum observed at $\approx$22~K reveals the presence of quantum fluctuations that suppress the onset of static magnetic order. Heat capacity measurements confirm the magnetic phase transition at $T_{\text{N}}$~=~6.32~K and also show a broad maximum at $\approx$10.6~K. 

Powder neutron diffraction reveals a commensurate magnetic structure with propagation vector $k$~=~(0,0,0) and a drastically suppressed ordered moment of 0.46~$\mu_\text{B}$/Cu$^{2+}$ at 2~K, compared to the free Cu$^{2+}$ moment of 1~$\mu_\text{B}$ again confirming the presence of strong quantum fluctuations in the ground state. Inelastic neutron scattering reveals that these fluctuations are due to the dominant antiferromagnetic J3 interaction which couples the spins into independent chains along the a-, b- and c-axes. The multispinon continuum, characteristic of the spin-1/2 HAFC is observed along the chain directions giving an intrachain interaction value J3~=~34.6(1)~K, while streaks of scattering are found perpendicular to the chains. The J1 and/or J2 interactions couple the chains together giving rise to the long-range magnetic order. Inspection of the magnetic structure shows that the spins form a 120$^{\circ}$ magnetic order about the triangles coupled by the J2 hyperkagome interaction suggesting that this interaction is antiferromagnetic. On the other hand the isolated triangles coupled by J1 show a strongly canted 120$^{\circ}$ structure that fully satisfies neither an antiferromagnetic nor a ferromagnetic interaction revealing that J1 is much weaker than J2.

Of the two isostructural compounds SrCuTe$_2$O$_6$ and PbCuTe$_2$O$_6$, BaCuTe$_2$O$_6$ appears to be most similar SrCuTe$_2$O$_6$ where the chain interaction J3 is dominant and antiferromagnetic and the ground state has long-range magnetic order. Their energy scales are also similar with J3$\approx$34~K for BaCuTe$_2$O$_6$ and $\approx$50~K for SrCuTe$_2$O$_6$. Nevertheless these two compounds have different magnetic structures, SrCuTe$_2$O$_6$ orders in the $\Gamma_1^1$ IR with a 120$^{\circ}$ spin arrangement around the J1 triangles suggesting that this interaction is antiferromagnetic and is responsible for coupling the chains together, while for BaCuTe$_2$O$_6$ which orders in the $\Gamma_2^1$ IR, J2 is antiferromagnetic and primarily responsible for the interchain coupling. The differences between BaCuTe$_2$O$_6$ and SrCuTe$_2$O$_6$ are probably related to the larger ionic radius of Ba$^{2+}$ compared to Sr$^{2+}$. In contrast PbCuTe$_2$O$_6$ is very different from both BaCuTe$_2$O$_6$ and SrCuTe$_2$O$_6$ even though Pb$^{2+}$ and Sr$^{2+}$ have the same ionic radius. In PbCuTe$_2$O$_6$ J1 and J2 are antiferromagnetic and of the same strength while J3 is significantly weaker, this difference might be related to the lone pair of Pb$^{2+}$. The crystal structure of the ACuTe2O6 family is complex and the magnetic interactions include super-super exchange paths, thus subtle differences in bond distances and angles, may cause significant changes in interaction strength.

To conclude BaCuTe$_2$O$_6$ is a new quantum magnet consisting of both frustrated and low dimensional interactions that result in long-range magnetic order co-existing with strong quantum fluctuations. Future studies will focus on obtaining a better understanding magnetic exchange paths from the low energy spin dynamics and $ab$ $initio$ calculations.

\begin{acknowledgments}
We thank Jean-S\'{e}bastien Caux for his calculation of the dynamical structure factor of the spin-1/2 Heisenberg antiferromagnetic chain. B.L. acknowledges the support of DFG through project B06 of SFB 1143 (ID 247310070). The powder synthesis, crystal growth and physical properties measurements took place at the Core Laboratory Quantum Materials, Helmholtz Zentrum Berlin f\"{u}r Materialien und Energie, Germany. We gratefully acknowledge the Science and Technology Facilities Council (STFC) for access to neutron beamtime at the LET and WISH ISIS facilities and also for the provision of sample preparation. We acknowledge Dr Nicola P. M. Casati from Laboratory for Synchrotron Radiation-Condensed Matter, Paul Scherrer Institut (PSI) for the synchrotron X-ray powder diffraction measurements on the Materials Science beamline (MS-X04SA).
\end{acknowledgments}

\appendix

\section{Structural determination}
\label{sec:StructDet}
\setcounter{figure}{0} \renewcommand{\thefigure}{A.\arabic{figure}} 
\setcounter{table}{0} \renewcommand{\thetable}{A.\arabic{table}} 
 
\begin{figure}
\centering
\begin{tabular}{c @{\qquad} c }
\includegraphics[width=0.8\columnwidth,keepaspectratio]{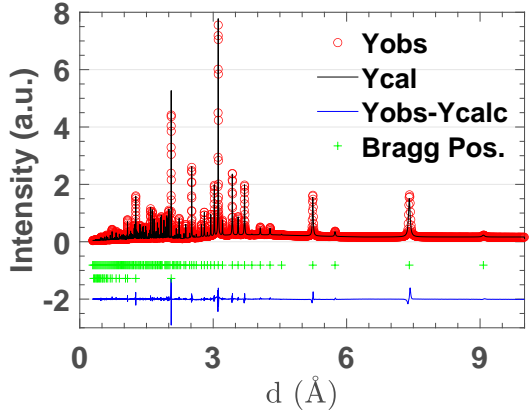} \\
\vspace*{0.3cm}
\small (a) \\
\includegraphics[width=0.8\columnwidth,keepaspectratio]{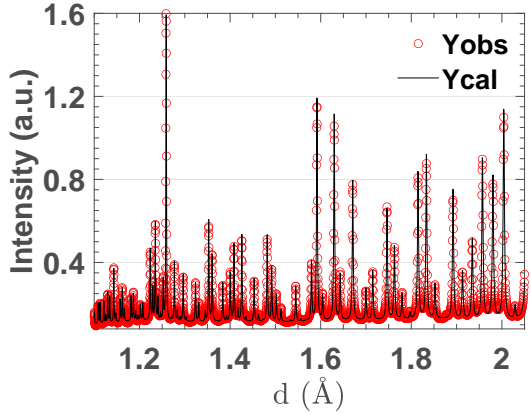}\\
\small (b) 
\end{tabular}
\caption{(a) Structural Rietveld refinement (black line) of the synchrotron X-ray diffraction pattern (red circles) collected for a crushed single crystal of BaCuTe$_2$O$_6$ at room temperature along with the difference between the fit and data (blue line). The green crosses indicate the Bragg peak positions for the two included phases, BaCuTe$_2$O$_6$ (top row) and diamond powder (bottom row). The refinement confirms the cubic space group (P$4_132 $) with lattice parameter a=12.8330(2)~\AA, achieved with $\chi^2$=15.0. (b) A small region of the diffraction pattern at low d-spacing showing the goodness of the refinement.} 
\label{XRD1}
\end{figure}

Figure~\ref{XRD1}(a) shows the room temperature high-resolution synchrotron X-ray diffraction pattern of BaCuTe$_2$O$_6$ measured on a crushed piece of single crystal. The diffraction pattern was refined within the cubic space group $P$4$_1$32. The wavelength of the synchrotron X-rays ($\lambda$=~0.4930(4)~\AA) and the instrument profile parameters were determined by refining the X-ray diffraction profiles of standard Si 640d and LaB$_6$ powders respectively which were collected with the same instrumental settings. From the Rietveld refinement the lattice parameter was found to be a=12.8330(2)~\AA. The goodness of the refinement is confirmed further by Fig.~\ref{XRD1}(b) which presents an expanded view focusing on the low $d$-spacing range. 

\begin{figure}
\centering
\begin{tabular}{c @{\qquad} c }
\includegraphics[width=0.8\columnwidth,keepaspectratio]{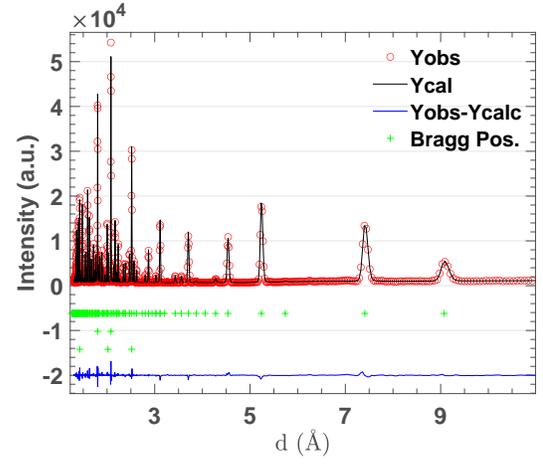} \\
\vspace*{0.3cm}
\small (a) \\
\vspace{-0.1cm}
\hspace*{-0.5cm} 
\includegraphics[width=0.85\columnwidth,keepaspectratio]{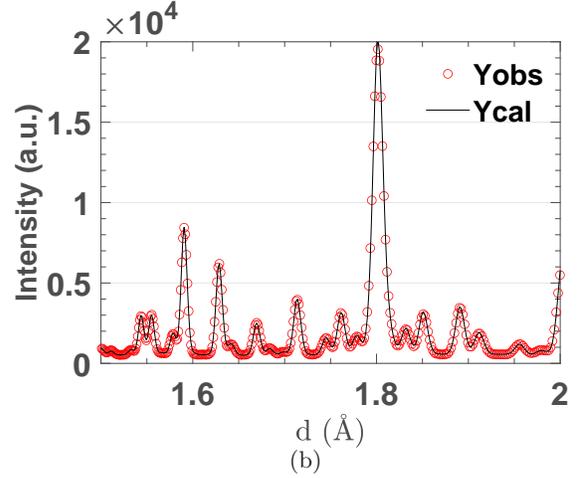} \\
\small (b) 
 \end{tabular}
\caption{(a) Rietveld refinement of the neutron powder diffraction pattern of the BaCuTe$_2$O$_6$ powder sample prepared by solid state reaction, collected at $T$=15~K on the SPODI diffractometer. The red open circles refer to the experimental data and the black and blue lines indicate the refined fit of the data and the difference between fit and experiment, respectively. The green crosses denote the Bragg peak positions for the three included phases, BaCuTe$_2$O$_6$, Cu container and the Al cryostat. The refinement confirms the cubic space group P$4_132$ with lattice parameter a=12.8328(1)~\AA. (b) A small region of the diffraction pattern at low d-spacing showing the goodness of the refinement. } 
\label{NeutronD}
\end{figure}

The neutron powder diffraction measurements were performed using the SPODI and WISH diffractometers. These measurements were performed on the powder sample of BaCuTe$_2$O$_6$ prepared by solid state reaction rather than the crushed single crystal. The sample was measured down to low temperatures so that any changes in crystal structure could be observed. All the data down to $T$=0.5~K, could be refined with the cubic space group  P4$_1$32 and no evidence for a structural phase transition was found. Fig.~\ref{NeutronD}(a) presents the powder diffraction pattern collected on SPODI at $T$=15~K along with the Rietveld refinement, the accuracy of the refinement is highlighted by a selected low $d$-spacing range, presented in Fig.~\ref{NeutronD}(b).

\section{Allowed magnetic peaks}
\label{sec:MagPeaks}
\setcounter{figure}{0} \renewcommand{\thefigure}{B.\arabic{figure}} 
\setcounter{table}{0} \renewcommand{\thetable}{B.\arabic{table}} 

\begin{figure}
\centering
\hspace*{-0.2cm} 
\begin{tabular}{c @{\qquad} c }
\includegraphics[width=1\linewidth,keepaspectratio]{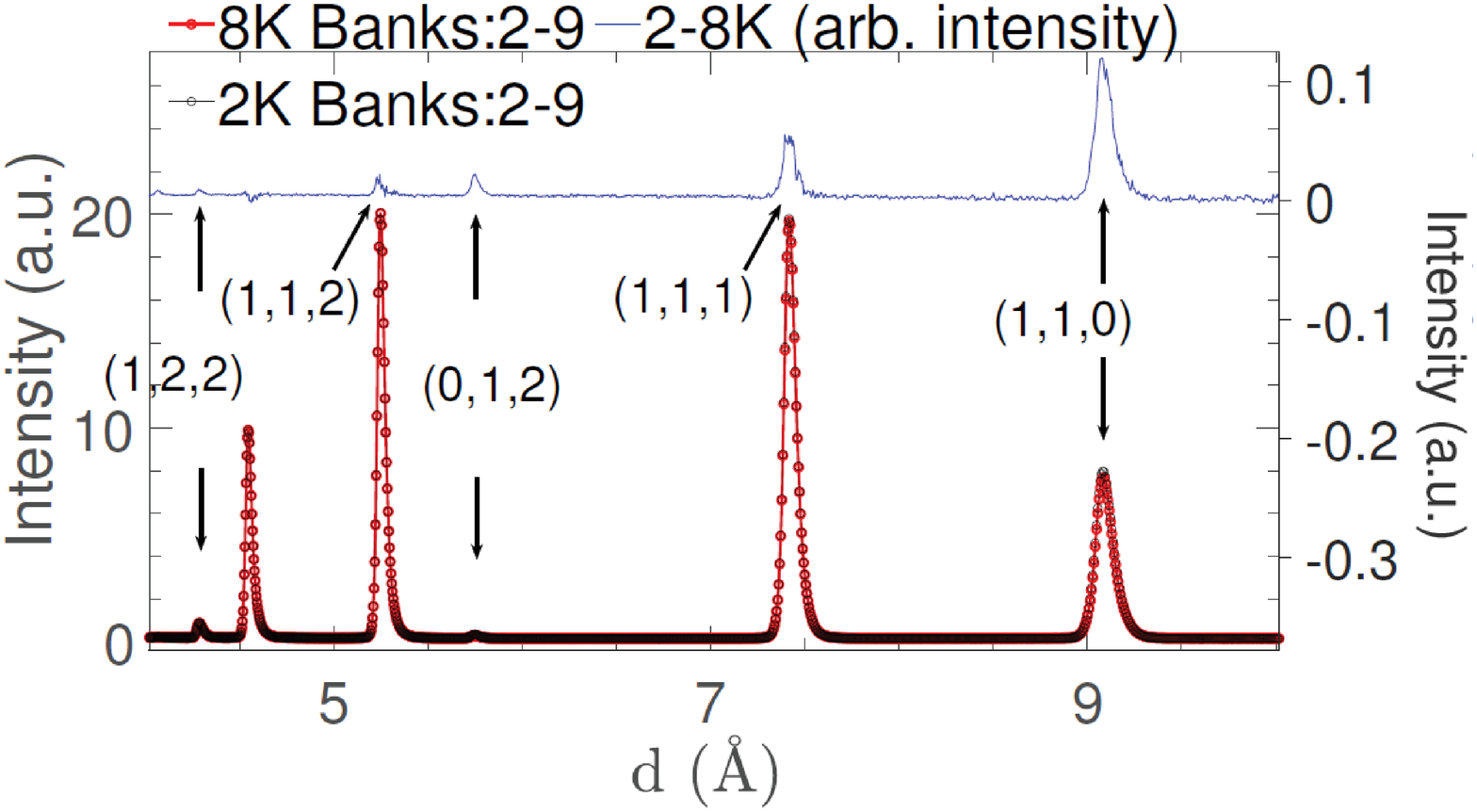} \\
\small (a) \\
\includegraphics[width=1\linewidth,keepaspectratio]{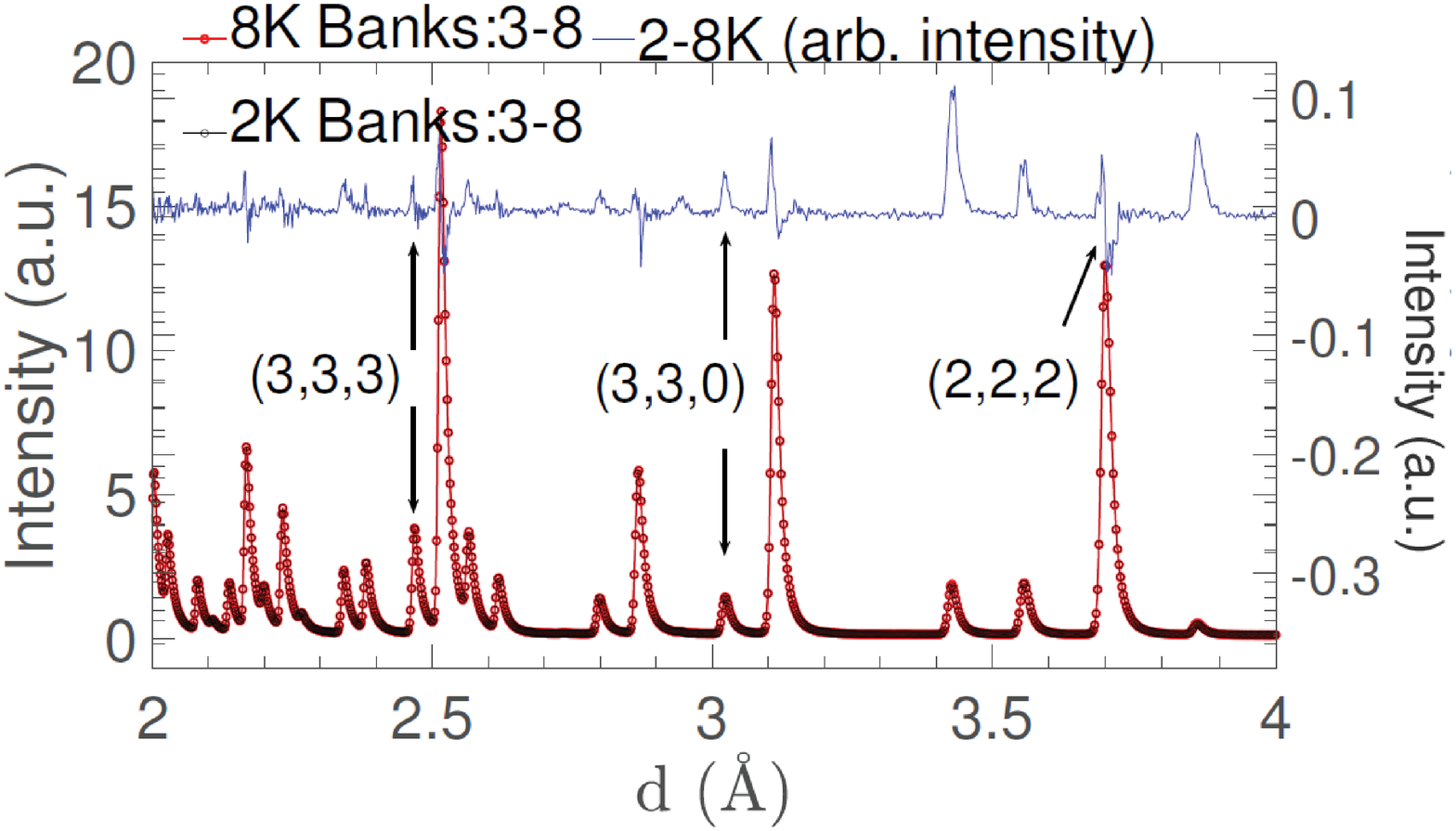}  \\
\small (b) \\
\includegraphics[width=1\linewidth,keepaspectratio]{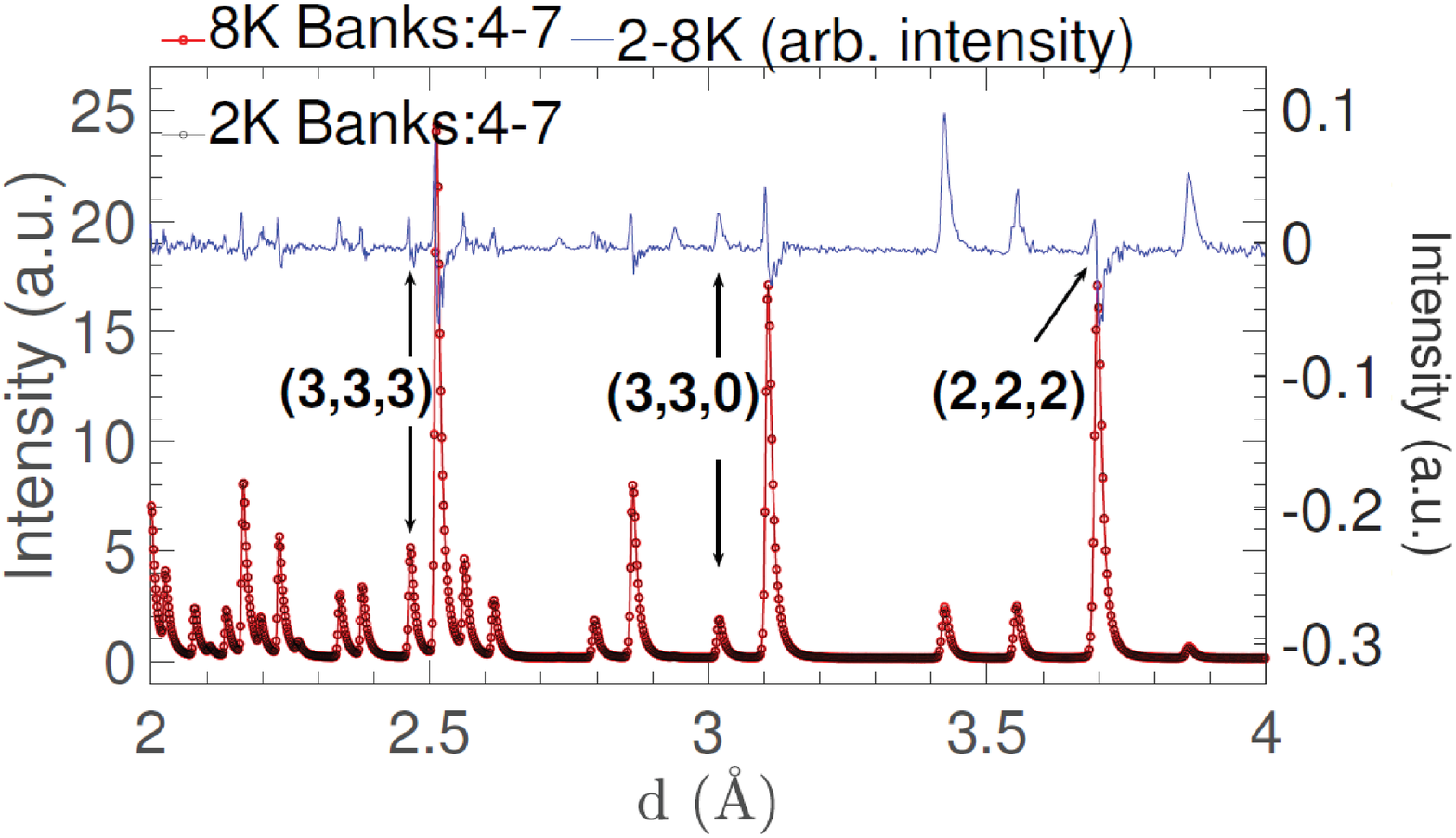}    \\
\small (c) 
 \end{tabular}
\caption{(a) Neutron powder diffraction data collected above and below the magnetic phase transition at $T$= 8~K (red points) and 2~K (black points) respectively on the WISH diffractometer plotted as a function of $d$-spacing. The difference pattern (dataset at 2~K with the 8~K dataset subtracted) is also plotted as shown by the blue symbols (the y-scale is shifted for clarity and shown on the right-hand side). The peaks are identified by their Miller indices. (a) shows the pattern at high $d$-spacing collected from detector banks 2 and 9 while the lower $d$-spacing range is shown by (b) detector banks 3 and 8, and (c) detector banks 4 and 7.}
\label{extraPeaks}
\end{figure}

Figure~\ref{extraPeaks} compares the neutron powder diffraction patterns collected above and below $T_{\text{N}}$ at $T$=8~K and $T$=2~K respectively for the various detector banks of the WISH diffractometer. The difference between the patterns (2~K dataset with 8~K dataset subtracted) is also presented in each case. The difference patterns show a series of peaks which are labelled by their Miller indices. Figure~\ref{extraPeaks}(a) shows the high $d$-spacing peaks (detector banks 2 and 9) and the difference pattern reveals the presence of (1,1,0), (1,1,1), (0,1,2) and (1,1,2).  It should be mentioned however that the (1,1,0) and (1,1,1) nuclear peaks are among those with the highest intensity and their apparent increase in intensity at low temperatures could be unreliable resulting from the subtraction of two large numbers rather than the presence of additional magnetic intensity.

In order to establish a consistent set of rules governing the magnetic peaks allowed in BaCuTe$_2$O$_6$, the remaining WISH detector banks, which focus on the lower $d$-spacings where the nuclear peaks that have weaker intensities, were also considered. Because these nuclear peaks are weaker, we anticipate that the result of subtracting the high from the low temperature datasets will be more reliable. Figure \ref{extraPeaks}(b) and (c) present the combined signal of detector banks 3 and 8 and banks 4 and 7 respectively. Both difference patterns clearly show the presence of the (3,3,0) reflection while the (2,2,2) and (3,3,3) reflections are absent. We conclude that the (H,H,0) reflections are allowed by the magnetic structure of BaCuTe$_2$O$_6$ but that the (H,H,H) reflections are absent, and therefore that the (1,1,1) peak observed in the difference pattern from detector banks 2 and 9 (Fig.~\ref{extraPeaks}(a)), is an artifact arising from the subtraction of the strong (1,1,1) nuclear reflection.

\section{Magnetic refinement using $1\times \Gamma_1^1$}
\label{sec:Gamma1}
\setcounter{figure}{0} \renewcommand{\thefigure}{C.\arabic{figure}} 
\setcounter{table}{0} \renewcommand{\thetable}{C.\arabic{table}} 

\begin{figure}
\centering
\begin{tabular}{c @{\qquad} c }
\vspace*{0.4cm}
\hspace*{0.2cm}
\includegraphics[width=0.83\linewidth,keepaspectratio]{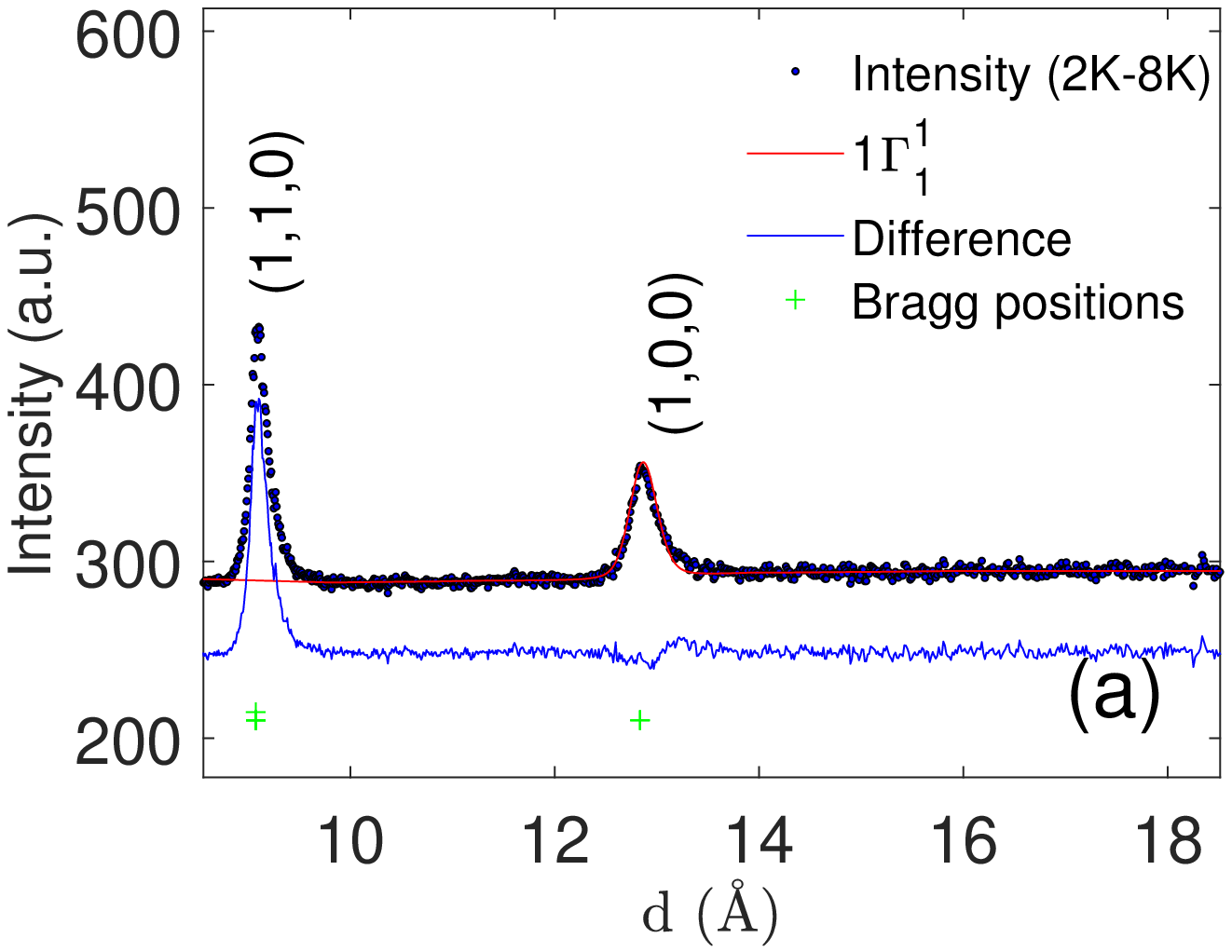}  \\         
\hspace*{0.2cm}
\vspace*{-0.4cm}
\includegraphics[width=0.83\linewidth,keepaspectratio]{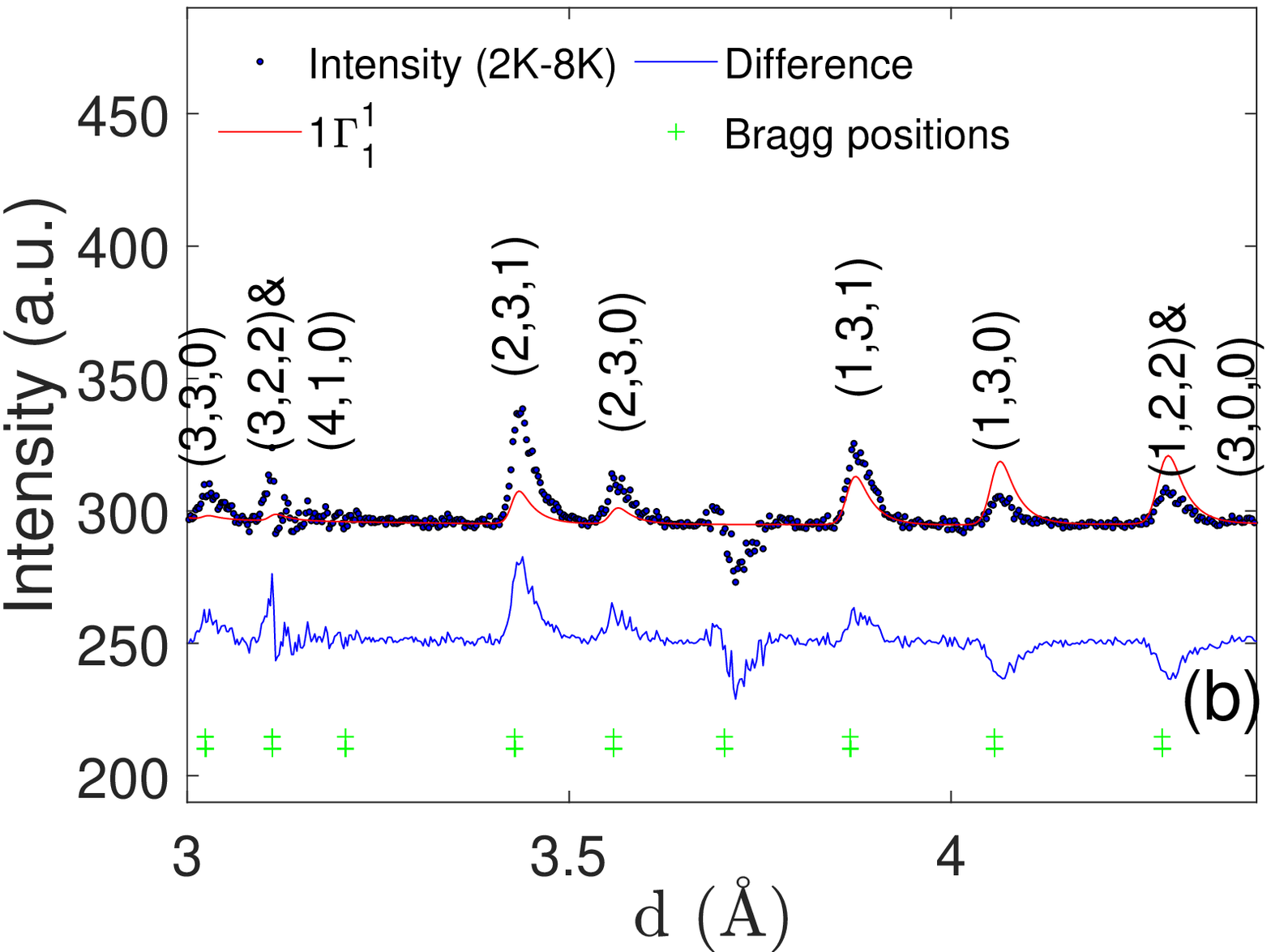}  \\
 \end{tabular} 
\vspace*{0.4cm} 
\caption{Rietveld refinement of the magnetic diffraction pattern collected on (a) detector banks 1 and 10 showing the high $d$-spacing peaks, and (b) detector banks 2 and 9 showing the low $d$-spacing peaks, of the WISH diffractometer for a powder sample of BaCuTe$_2$O$_6$. The black circles give the result of subtracting the 8~K dataset from the 2~K dataset, the red line gives the refined fit of the data assuming the IR $1\times \Gamma_1^1$ ($\chi^2$~=~8.5) while the blue line gives the difference between theory and experiment. 
}
\label{bcto_wish_Fit_G1}
\end{figure}

Since the magnetic structure of SrCuTe$_2$O$_6$ which is isostructural to BaCuTe$_2$O$_6$ orders within the $1\times \Gamma_1^1$ IR \cite{Chillal2020,Saeaun2020}, this IR was used to refine BaCuTe$_2$O$_6$ for comparison with the $2\times \Gamma_2^1$ IR refinement given in Fig.~\ref{bcto_wish_Fit}.
Figures~\ref{bcto_wish_Fit_G1}(a) and (b) present the Rietvelt refinement (red line) of the magnetic diffraction pattern (black circles) collected on detector banks 1 and 10 showing the high $d$-spacing peaks, and detector banks 2 and 9 showing the low $d$-spacing peaks respectively. $1\times \Gamma_1^1$ IR forbids both the (H,H,0) and (H,H,H) reflections thus it fails to reproduce the intensity at the (1,1,0) peak and generally gives a much poorer fit over all $d$-spacing ranges with a $\chi^2$ value of 8.5 compared to 3.1 for the $2\times \Gamma_2^1$ IR.

\end{document}